\documentclass[fleqn,usenatbib]{mnras}

\usepackage{preface}
\usepackage{import}
\usepackage{float}
\usepackage{soul}
\usepackage{xcolor}

\usepackage{newtxtext,newtxmath}

\title[Background Systematics of the GCE in M31]{The Andromeda Gamma-Ray Excess: Background Systematics of the Millisecond Pulsars and Dark Matter Interpretations}

\author[Zimmer et al.]{
\parbox[t]{\textwidth}{
Fabian Zimmer,$^{1}$\thanks{f.zimmer@uva.nl}
Oscar Macias,$^{1,2}$\thanks{o.a.maciasramirez@uva.nl}
Shin'ichiro Ando$^{1,2,3}$,
Roland M. Crocker$^{4}$,
Shunsaku Horiuchi$^{5,3}$
\\}
\\
$^{1}$GRAPPA $-$ Gravitational and Astroparticle Physics Amsterdam, University of Amsterdam, Science Park 904, 1098 XH Amsterdam, The Netherlands\\
$^{2}$Institute for Theoretical Physics Amsterdam and Delta Institute for Theoretical Physics, University of Amsterdam, Science Park 904,\\ 1098 XH Amsterdam, The Netherlands.\\ 
$^{3}$Kavli IPMU (WPI), UTIAS, The University of Tokyo, Kashiwa, Chiba 277-8583, Japan\\
$^{4}$Research School of Astronomy and Astrophysics, Australian National University, Canberra 2611, A.C.T., Australia\\
$^{5}$Center for Neutrino Physics, Department of Physics, Virginia Tech, Blacksburg, VA 24061, USA\\
}

\date{Accepted XXX. Received YYY; in original form ZZZ}

\pubyear{2022}

\begin{document}
\label{firstpage}
\pagerange{\pageref{firstpage}--\pageref{lastpage}}
\maketitle

\begin{abstract}
Since the discovery of an excess in gamma rays in the direction of M31, its cause has been unclear. Published interpretations focus on a dark matter or stellar related origin. Studies of a similar excess in the Milky Way center motivate a correlation of the spatial morphology of the signal with the distribution of stellar mass in M31. However, a robust determination of the best theory for the observed excess emission is very challenging due to large uncertainties in the astrophysical gamma-ray foreground model. Here we perform a spectro-morphological analysis of the M31 gamma-ray excess using state-of-the-art templates for the distribution of stellar mass in M31 and novel astrophysical foreground models for its sky region. We construct maps for the old stellar populations of M31 based on observational data from the \textsc{PAndAS} survey and carefully remove the foreground stars. We also produce improved astrophysical foreground models by using novel image inpainting techniques based on machine learning methods. We find that our stellar maps, taken as a proxy for the location of a putative population of millisecond pulsars in the bulge of M31, reach a statistical significance of $5.4\sigma$, making them as strongly favoured as the simple phenomenological models usually considered in the literature, e.g., a disk-like template with uniform brightness. Our detection of the stellar templates is robust to generous variations of the astrophysical foreground model. Once the stellar templates are included in the astrophysical model, we show that the dark matter annihilation interpretation of the signal is unwarranted. Using the results of a binary population synthesis model we demonstrate that a population of about one million unresolved MSPs could naturally explain the observed gamma-ray luminosity per stellar mass, energy spectrum, and stellar bulge-to-disk flux ratio.

\end{abstract}


\section{Introduction}
\label{sec:intro}

Due to its proximity and mass, the center of the Milky Way (MW) is expected to be the brightest source of dark matter (DM) annihilation in the sky~\citep[e.g.,][]{Bertone:2005,Charles:2016pgz}. However, our view of the MW halo is obscured by large amounts of uncertain interstellar material. It is thus vital to carry out complementary searches for DM emission in regions with differing astrophysical uncertainties.

The Andromeda galaxy (M31), located at a somewhat low Galactic latitude $(l,b)=(121.17^\circ, -21.57^\circ)$, suffers lower foreground extinction than the MW center. Moreover, due to the large inclination angle (${\sim}77.5^\circ$) of the plane of M31 with the line of sight \citep{Tamm:2012}, we can observe the M31's halo almost completely unobstructed by its stellar and gaseous disk. A wealth of recent observations serve to elucidate key differences between the MW and M31, which are thought to be due to their different accretion histories ~\citep{Kormendy:2013nda}. M31 has a stellar mass that is similar or possibly even larger than that of the MW~\citep[e.g.,][]{Watkins:2010,Dias:2014}. Furthermore, the M31's bulge is a factor of 5 to 6 times more massive than the compact bulge/bar system in the MW~\citep{Tamm:2012, Licquia:2015}, its supermassive black hole is about 50 times more massive than Sgr A$^\star$~\citep{Gerhard:2016}, and the star formation rate (SFR) in M31 is a factor of about 10 times lower than in the MW~\citep{Ford:2013}. Altogether, M31 constitutes not only an excellent target for searches of DM emission~\citep[e.g.,][]{Lisanti:2017qlb}, but also a unique stepping stone in our efforts to understand the high-energy astrophysics of spiral galaxies.

Interestingly, there has been a recent discovery of an excess in gamma-rays coming from the inner region of M31 \citep{Ackermann:2017nya}, which appears to have similar characteristics to those of the longstanding so-called Galactic Center Excess (GCE) in the MW~\citep[e.g.,][]{Hooper:2010mq,Abazajian:2012pn,Gordon:2013vta, Macias:2013vya,Daylan:2014rsa,Calore:2014nla,TheFermi-LAT:2015kwa,TheFermi-LAT:2017vmf}. In particular, using almost seven years of \textit{Fermi}-LAT data, \cite{Ackermann:2017nya} detected extended diffuse gamma-ray emission from the inner $\sim 0.4^\circ$ (or $\sim 5$ kpc) of the M31's halo, at the $4\sigma$ statistical level. Its spatial morphology was found to be compatible with a uniform/Gaussian disk, and its spectrum was not well constrained---either a simple power-law or a bump-like spectrum could give an acceptable fit to the data. Importantly,  they obtained that M31's emission is not correlated with the distribution of interstellar gas nor regions of star formation activity, which are both mostly located in a large ring-like structure at $\sim 10$ kpc from its center. However, gamma-ray emission from the gaseous disk was not strongly ruled out and might even be present up to the 50\% level of the measured flux~\citep{Ackermann:2017nya}. In addition, a later in-depth large field study~\citep{Karwin:2019jpy} of the outer halo of M31 found evidence for emission within the $\sim 120-200$ kpc (or $\sim 8^\circ - 12^\circ$) of its center.

Recent theoretical studies have proposed different alternative explanations for the inner galaxy M31 excess: cosmic ray (CR) models~\citep[e.g.,][]{McDaniel:2019niq,2021PhRvD.104l3016D}, unresolved millisecond pulsars (MSPs)~\citep[e.g.,][]{Eckner:2017oul,Fragione:2018jjg}, and DM emission models~\citep[e.g.,][]{Karwin:2020tjw,Burns:2020mti,Chan:2021dyy}. As for the excess in the outer halo of M31, \cite{2021ApJ...914..135R} posited some well motivated CR models.

Since the M31 signal is only an \enquote{excess} with respect to our current understanding of the diffuse gamma-ray background, it is important to investigate the impact that diffuse mismodeling has on the characteristics of the signal. Very recent studies~\citep[e.g.,][]{PhysRevD.103.083023,DiMauro:2019frs} have used either \texttt{SkyFACT}~\citep{Storm:2017arh}, or alternative diffuse emission models built with GALPROP~\citep{Strong:1998pw} for the evaluation of these uncertainties. In the present study, we perform a reanalysis of this data with a novel approach; we construct tailor-made foreground gas map models for the M31 region which are expected to be less affected by biases. \cite{Karwin:2020tjw} highlighted that the foreground interstellar gas maps used in the identification of the M31 excess might contain a fraction of the interstellar gas that should belong to the M31 galaxy. Here, we excise M31 from the foreground interstellar gas models, followed by using various inpainting techniques (e.g. the methods ``Nearest-Neighbour" and ``\texttt{SMILE}") for reconstruction or processing all the information of the image by a neural network (\enquote{Deep-Prior} method) to inpaint the excised region. In this way we restore the actual spatial distribution of the interstellar gas in the foreground as best as possible. The resulting alternative maps are then employed in our reassessment of the characteristics of the M31 excess. 

Studies of the GCE~\citep{Macias:2016nev, Bartels2018, Macias:2019omb, Coleman:2019kax, Abazajian:2020tww, 2022arXiv220311626P} have demonstrated that the two most compelling explanations (astrophysical or DM emission) can be distinguished based on their spatial morphologies. Indeed, such studies have consistently found that the morphology of the GCE traces the stars of the inner Galaxy better than it traces the distribution expected for DM (however, see \cite{DiMauro:2021raz} for opposite conclusions). Motivated by the very promising morphological results obtained in the GCE, here we attempt to separate the stellar from the DM hypotheses using a traditional fitting regression technique. For this, we construct empirical models of the stellar distribution in M31 using data from the \textsc{PAndAS} survey, which has been mapping out the environment of M31 for more than a decade. We construct stellar maps for the disk based on this data and using the stellar contamination model of \citet{Martin_2013}, which effectively removes the stars residing in the Milky Way. Since, unfortunately, this data is unreliable for the inner most  regions of M31, we use instead data from WISE \citep{Wright_2010} to construct the bulge component of our stellar templates.

The paper is outlined as follows. In Sec. \ref{sec:data} we describe the data used in this work. In Sec. \ref{sec:methods} we present the methods used to construct the diffuse emission components and the various models used in our analysis. In Sec. \ref{sec:results} we present the results of the gamma-ray analyses and discuss their implications in Sec. \ref{sec:conclusions}.

\section{Gamma-ray Observations}
\label{sec:data}

In this work we use data from the {\it Fermi} Large Area Telescope ({\it Fermi}-LAT), covering 10 years of observations (see Table~\ref{tab:selection_criteria}). We use exactly the same data selection cuts assumed in the construction of the 4FGL-DR2 catalog \citep{Fermi-LAT:2019yla, 2020arXiv200511208B} to avoid potential biases induced by new point sources (not present in the 4FGL-DR2 catalog) which would likely appear in our region of interest (ROI) if we were using a longer observation time.

We restrict ourselves to data from a 14 by 14 degree region in the sky, centered on the SIMBAD coordinates of M31, i.e. $(l,b)=(121.17^\circ, -21.57^\circ)$, and spanning an energy range of 500 MeV to 100 GeV. 
We are most interested in the spatial morphology of the signal, since we want to determine in which part of the M31 galaxy the signal is strongest. We therefore chose a lower limit of 500 MeV to preserve as much angular information as possible, without too much of a loss of photon statistics.
Previous studies have gone lower to 300 MeV \citep{PhysRevD.103.083023} or even as low as 100 MeV \citep{DiMauro:2019frs}. These studies have used the \textit{SOURCE} event class data, while we restrict ourselves to the \textit{CLEAN} event class data. This latter dataset benefits from a lower background rate than the \textit{SOURCE} event class. 

We make use of the \textit{Fermi} \texttt{ScienceTools} version 2.0.0 software package to further reduce the raw data. Since the Pass 8 data release, it is possible to select data based on the reliability of the reconstructed direction based on the instrument's point spread function (PSF). We want to use photons from the PSF3 quartile only (the most reliable in reconstructed direction), which would benefit our morphological analysis the most. However, due to our bin-by-bin approach, higher energy bins lack statistical power if we do not include other PSF quartiles causing  reconstructed spectra to fluctuate. Therefore, we restricted our analysis to the PSF3 quartile only for the first three energy bins, where the counts were high enough, and we used all PSF quartiles combined for the rest of the bins. 

All the technical details of the selection filters applied to the data by the Fermitools are summarized in Table \ref{tab:selection_criteria}. For more information regarding the criteria, we refer the reader to the official {\it Fermi} website\footnote{https://fermi.gsfc.nasa.gov/ssc/data/analysis/documentation/Cicerone}.

\begin{table}[h]
\caption{{\it Fermi}-LAT data selection criteria used for our analysis (applied to the data with the \textit{Fermi} \texttt{ScienceTools}).}
\begin{tabular}{ c c }
\hline
\hline
Category & Selection criteria \\ 
\hline
Observation period & Aug 4, 2008
to Aug 2, 2018 \\
Mission elapsed time (s) & 239557417 to 554929985 \\
Energy range (GeV) & 0.5 to 100 \\
Energy binning & 10 bins (log spaced) \\
Event class & Pass 8 CLEAN \\
Event type & PSF3 for [0.5,2.5] GeV, \\ & all PSF for [2.5,100] GeV \\
IRFS & P8R3 CLEAN V3 \\
$z_{max}$ & $90^{\circ}$ \\
Filters & (DATA\textunderscore QUAL>0) $\&$ (LAT\textunderscore CONFIG=1) \\
Region of interest & $14^{\circ} \times 14^{\circ}$ at $(l,b) = (121.17^\circ, -21.57^\circ)$ \\
Pixel resolution & $0.1^{\circ}$ \\
Pixel binning & $140 \times 140$ pixels \\
Sky projection & Cartesian "CAR" \\
\hline
\hline
\end{tabular}
\label{tab:selection_criteria}
\end{table}

\section{Methods}
\label{sec:methods}

In this section we describe the construction of tailor-made Galactic diffuse emission models for our ROI. One of the biggest challenges in creating an appropriate foreground model for the M31 region, is that it is very difficult to disentangle the foreground hydrogen gas that belongs to the MW from the actual gas in M31. The standard approach followed by the {\it Fermi} team consists of removing M31 from the hydrogen gas maps, based on cuts in the $(l,b,v_{LSR})$ data space with $v_{LSR}$ being the local standard of rest velocity. After it has been excised, they reconstruct the foreground hydrogen maps based on the techniques described in \cite{ELAD2005340,Acero:2016qlg}. Although this is a reasonable analysis choice, it is expected to introduce a bias in the gamma-ray properties of M31. In particular, it has been pointed out~\citep{Karwin:2019jpy} that the hydrogen maps used in the official foreground \textit{Fermi} Galactic diffuse emission map might be holding some hydrogen gas which belongs to M31. In this work, we aim to construct more suitable Galactic foreground maps by using novel inpainting algorithms based on neural networks and deep learning to construct M31-free Galactic foreground templates. We also evaluate the uncertainties associated to this method by rerunning our gamma-ray pipeline with all the resulting alternative foreground maps.

\subsection{Standard \textit{Fermi} Galactic Diffuse Emission Model}
\label{sec:fermi_bkg}

One of the models we considered in this work is the so-called  \verb|gll_iem_v07| model. This is the standard Galactic diffuse emission model recommended for analyses of point-like and small-sized extended sources, constructed by the {\it Fermi} collaboration---henceforth simply called the {\it Fermi} background.   

The {\it Fermi} background is made up of a linear combination of templates, each one responsible for capturing the contribution of a physical mechanism to the total gamma-ray emission. Some of the largest contributions come from the decay of energetic neutral pions, and electron bremsstrahlung radiation. Since these two components are spatially correlated with gas, they were phenomenologically modeled using atomic and molecular hydrogen maps constructed from radio observations~\citep{Acero:2016qlg}.

Another big contribution to the {\it Fermi} background comes from the inverse-Compton scattering process. This component is very difficult to model correctly as it depends on many factors, which include solving the cosmic-ray transport equation. The {\it Fermi} team included inverse-Compton emission templates generated with the numerical code \texttt{GALPROP} \citep{Strong:1998pw}.

The procedure followed by the {\it Fermi} team was then to fit these components---in addition to other empirical templates accounting for positive and negative residuals in the data---to the observed all-sky gamma-ray data using a template fitting approach. Once the best-fit fluxes for each template were determined, they combined all these in one compactified energy-dependent spatial template. This constitutes the \verb|gll_iem_v07| model.

\subsection{Alternative Foreground/Background Diffuse Emission Models}
\label{sec:selfmade_bkgs}

\subsubsection{Gas-correlated gamma-ray Component}
As in the {\it Fermi} background, the gas-correlated emission is modeled by assuming that the gamma-ray intensity is proportional to the interstellar gas column density---mainly atomic (H$I$) and molecular (H2) hydrogen. We use the high resolution H$I$4PI survey \citep{2016A&A...594A.116H} as in the {\it Fermi} background as our first option. As a second option we use hydrodynamical gas and dust maps, i.e., H$I$ and H2 gas column density templates developed in \cite{Macias:2016nev}, which were already used to show that the GCE can be explained by an unresolved MSP population rather than annihilating DM \citep{Macias:2019omb,Abazajian:2020tww}. These templates were originally designed for analyses related to the MW Galactic center and for this purpose split into four galactocentric rings. Only the rings 3 and 4 of the H$I$ component contribute to the M31 region, whereas none do for the H2 component. 

Similarly to the {\it Fermi} collaboration, we also excise M31 from the gas maps where present, but use more sophisticated inpainting algorithms to fill in the regions of missing data. Two methods for inpainting are employed. The first one makes use of deep convolutional neural networks to restore the image. It was developed in \cite{Puglisi:2020deh}, where they successfully recovered the necessary statistical properties the image had before the reconstruction process. The authors tested it in the context of synchrotron and dust polarized emission, which represent the Galactic contamination in cosmic microwave background measurements. Their code was made publicly available as the \textbf{P}ython \textbf{I}npainter for \textbf{C}osmological and \textbf{AS}trophysical \textbf{SO}urces (\texttt{PICASSO}) package. We successfully employed two methods of their code for our purposes: The \enquote{Nearest-Neighbour} algorithm based on \cite{Bucher:2015ura} and the \enquote{Deep-Prior} method based on \cite{Ulyanov_2020}.

The other inpainting algorithm we use is from \cite{Li2014Smile}, which we call the \texttt{SMILE} inpainting algorithm. It is based on solving the Laplace equation for each missing pixel by using the values of the surrounding pixels\footnote{For an example of a similar method implemented in Mathematica see https://community.wolfram.com/groups/-/m/t/873396}. Such an approach has been successfully used by \citet{Macias:2019omb} to inpaint masked point-like sources near the center of the Milky Way. 

\begin{figure*}
    \centering
    \includegraphics[width=0.75\textwidth]{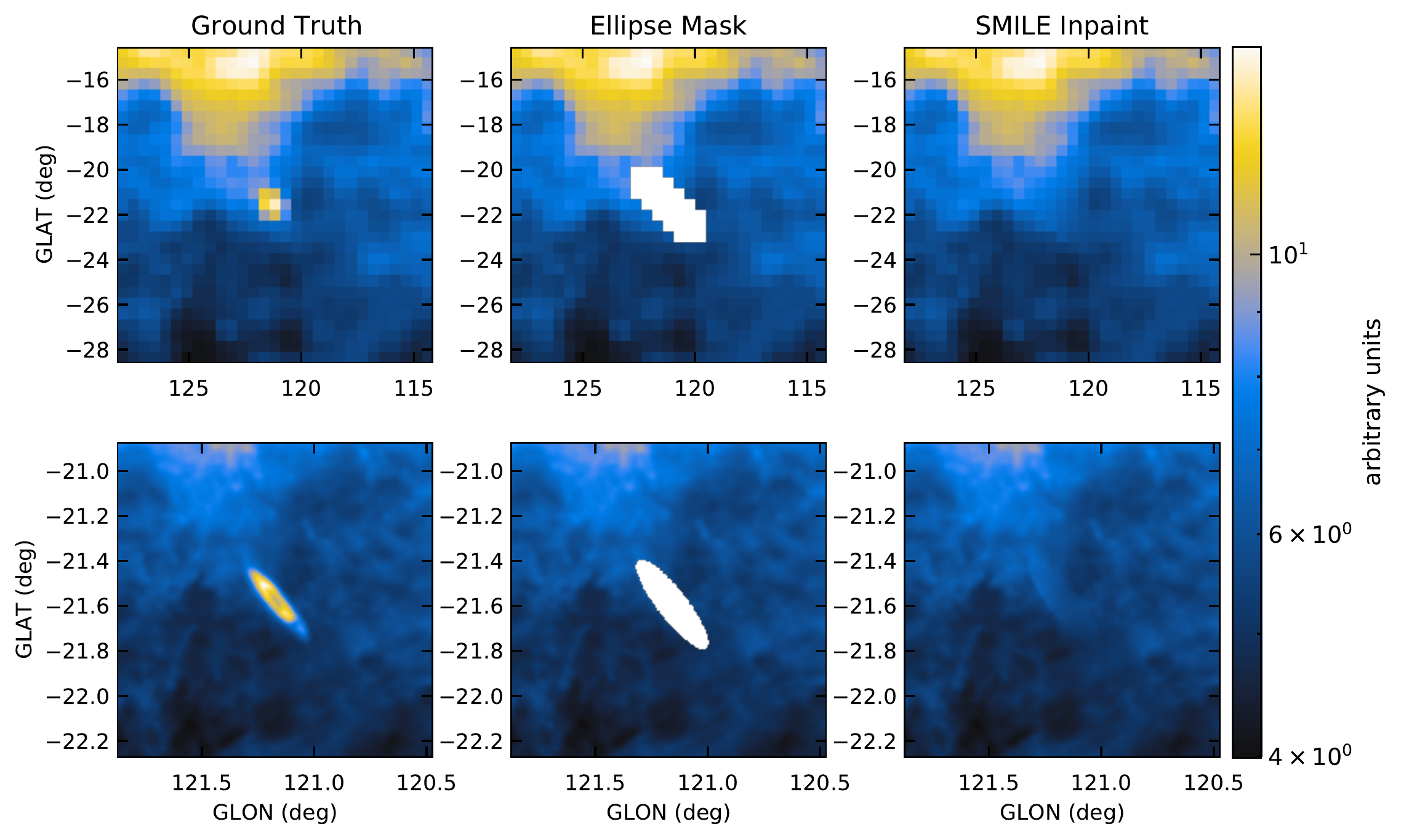}
    \caption{Visualization of the inpainting process when using the \texttt{SMILE} algorithm. From left to right, the three columns show the original hydrogen template, the masked region and the inpainted final version. The upper and lower rows show the H$I$ component of \protect\cite{Macias:2016nev} (as described in the text) and the H$I$4PI map, respectively.}
    \label{fig:smile_inpaint_visuals}
\end{figure*}

\begin{figure*}
    \centering
    \includegraphics[width=0.9\textwidth]{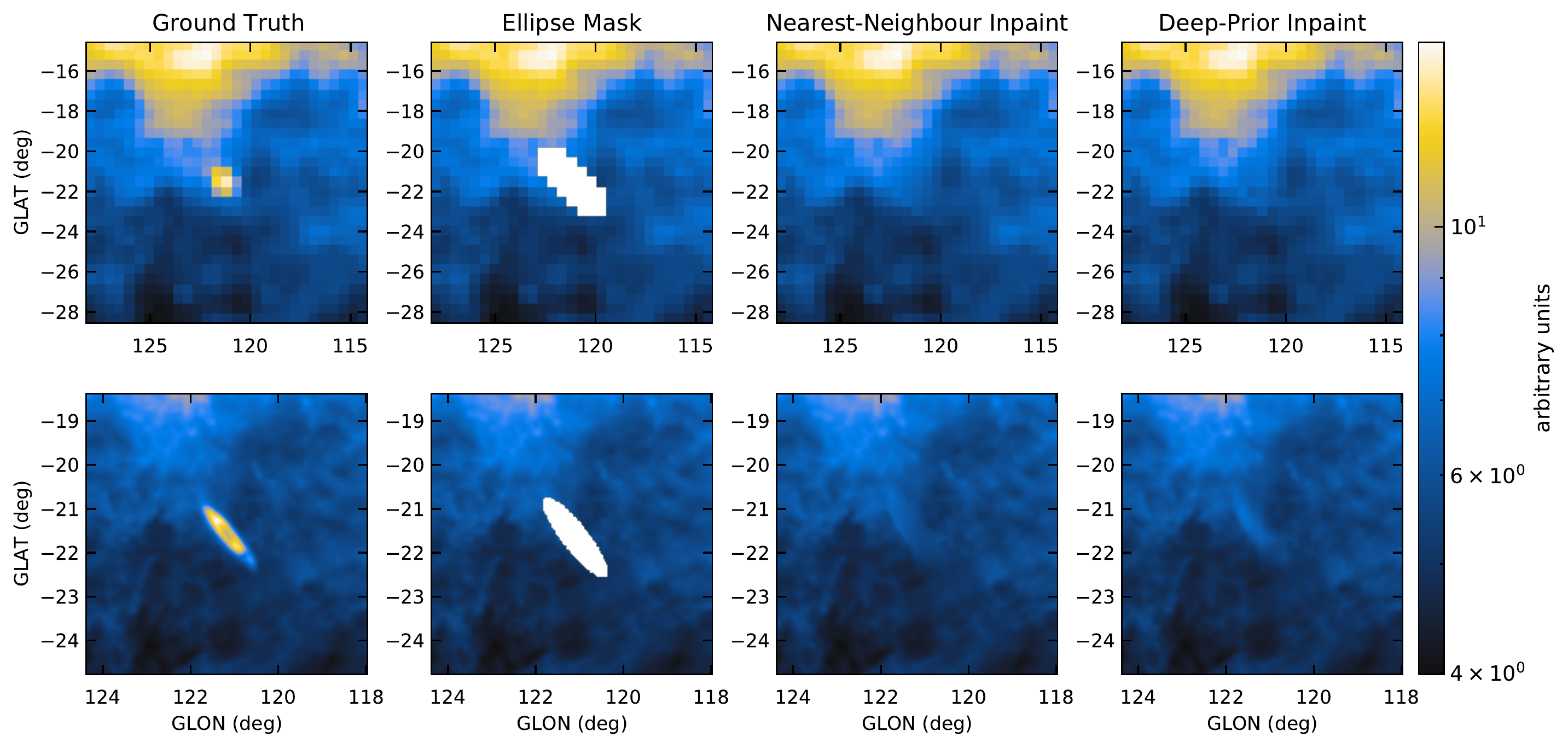}
    \caption{Visualization of the inpainting process when using the \texttt{PICASSO} algorithms. From left to right, the four columns show the original hydrogen template, the masked region and the final inpainted versions when using the Nearest-Neighbour or the Deep-Prior algorithm. The upper and lower rows show the H$I$ component of \protect\cite{Macias:2016nev} (as described in the text) and the H$I$4PI map respectively.}
    \label{fig:picasso_inpaint_visuals}
\end{figure*}

Visual representations of the inpainting procedures when using either \texttt{SMILE} or \texttt{PICASSO} are shown in Fig. \ref{fig:smile_inpaint_visuals} and Fig. \ref{fig:picasso_inpaint_visuals}, respectively. The first column of panels in each figure show the original map, where M31 is present. The second column depicts the extension of the masked region, which we based on cuts described in \cite{Ackermann:2017nya}, which were shown to effectively remove the majority of the disk of M31 from the data. The region was then inpainted with the aforementioned tools and we chose to use these algorithms over the one previously used by the {\it Fermi} collaboration due to their simpler implementation and to assess the impact on the properties of the M31 excess due to systematic uncertainties in the gamma-ray background model. These inpainted, M31-free versions can be seen in the last columns. In both figures, the lower row corresponds to the H$I$4PI map and the upper row to the atomic hydrogen column density map used in \cite{Macias:2016nev}. Note that the latter was derived from the 21 cm LAB survey~\citep{Kalberla:2005ts}, which has much lower angular resolution than the newer H$I$4PI survey.

\subsubsection{Inverse-Compton Component}
The second largest contribution to the diffuse gamma-ray sky comes from the inverse-Compton radiation produced by the photon upscattering by energetic electrons. All of the models for this inverse-Compton component we employ in this work are generated with the Galactic cosmic-ray propagation code \texttt{GALPROP} \citep{Strong:1998pw}. 

We used the three models from \cite{Ackermann:2014usa} labeled Foreground Model A, B and C. This study was interested in investigating how their fit results were affected by the specific type of foreground model chosen. These three models are thought to encompass a very wide range of systematic uncertainties associated with the inverse-Compton component. Model A uses a distribution of cosmic-ray sources based on \cite{Lorimer:2006qs} and standard choices for the propagation parameter setup. For Model B they include an additional population of sources for electrons near the Galactic center, which made this model a better fit according to their analysis. Also, Model B predicts an enhanced gamma-ray emission at high latitudes. This makes it highly relevant for out study as our ROI is relatively far from the plane. The diffusion and reacceleration parameters of Model C vary with Galactocentric radius throughout the Galaxy, as opposed to the constant diffusion coefficient assumed in the two other models. 

Additionally, we use the inverse-Compton model of \cite{Abazajian:2020tww}, which is split into six galactocentric rings. Only their local and outermost rings contribute to the M31 sky region and are used in our analyses.

By combining templates for these two major components and different inpainting algorithms we arrive at a total of 2 (H1 components) $\times$ 3 (Inpainting Methods) $\times$ 4 (Inverse-Compton Models) = 24 combinations. By testing models over this kind of variety of background models, we can classify the obtained significances of detection against the systematic uncertainties that come with the modelling of the Galactic diffuse gamma-ray emission. In addition to the gas-correlated and inverse-Compton templates, we use the same components as in the {\it Fermi} background for the isotropic gamma-ray background, emission from the sun and moon, and the same 4FGL-DR2 source catalog.

\subsection{Phenomenological M31 Models} 
\label{sec:pheno_models}

All the phenomenological models we used are integrated in \texttt{Fermitools}. We tested whether the signal was point-like or spatially extended. Specifically, we used disks of different radii and uniform brightness or a Gaussian disk of different variance. In most of the studies related to this phenomenon, investigators have included similar templates in their analyses, from the discovery of the excess \citep{Ackermann:2017nya} to both the previously mentioned studies closely related to our work \citep{DiMauro:2019frs,PhysRevD.103.083023}. The purpose of these simple constructions is to test whether the data prefers an extension beyond a point-like emission, but these extended templates are likely unphysical in the sense that real emission will fail to be as symmetric and clean as they are. Arguably, better suited models for this type of analysis are based on multi-wavelength observations from space telescopes and ground detectors.

\subsection{Interstellar Gas and Dust Models of M31} 
\label{sec:gas_and_dust_models}

One of the difficulties in astronomical studies like ours, where we want to isolate specific astrophysical gamma-ray sources, is the interference of emission from gas and dust clouds. They can either be part of the Andromeda system or reside inside the Milky Way halo, extending across the line of sight between us and M31. There has even been evidence for hydrogen structures extending between the two galaxies in interstellar space \citep{Lockman_2012}. 

There have been studies dedicated to mapping out the gas and dust contents of the Andromeda galaxy. A deep wide-field H$I$ imaging survey has been done in \cite{Braun_2009} with the \textit{Westerbork Synthesis Radio Telescope}. We also include two different versions of this hydrogen map labeled by BraunV2 and BraunV3, which include corrections for opacity effects. An intensive study about Andromeda's dust was done in \cite{Draine:2013mfa} using observations from the \textit{Spitzer Space Telescope} and \textit{Herschel Space Observatory}. We use these three maps\footnote{All data products related to these studies can be downloaded from https://www.astro.princeton.edu/~draine/m31dust/m31dust.html.} to account for unmodeled residual gas and dust emission belonging to M31.

\subsection{Construction of the Stellar Density Templates}
\label{sec:Starmaps}

One of the most important undertakings of this work is to test whether the data prefer an explanation involving MSPs or DM. To that end, we constructed stellar density maps consisting of old red giants (from now on simply called stellar maps), which we use as proxy for the spatial distribution of MSPs given that they are both old stellar populations.

We will begin by describing the construction of our stellar maps, for which we adapted the procedure of \cite{Martin_2013}. The data used to construct our stellar maps stems from the Pan-Andromeda Archaeological Survey (PAndAS) \citep{Ibata:2013aua, McConnachie_2018}. The data is publicly available at the PAndAS archive\footnote{Queries can be made \href{http://www.cadc-ccda.hia-iha.nrc-cnrc.gc.ca/en/community/pandas/query.html}{here}.}. 

First, we selected all stars within a 150 kpc radius from the PAndAS survey, centered on the SIMBAD coordinates of the center of M31. These stars are then de-reddened to correct for extinction by using the $E (B-V)$ map from \cite{Schlegel:1997yv} and the corrected color magnitudes are then obtained with

\begin{align}
\begin{split}
    g_0 &= g - 3.793 E(B-V) \\
    i_0 &= i - 2.086 E(B-V).
\end{split}
\end{align}

\begin{figure}
    \centering
    \includegraphics[width=\linewidth]{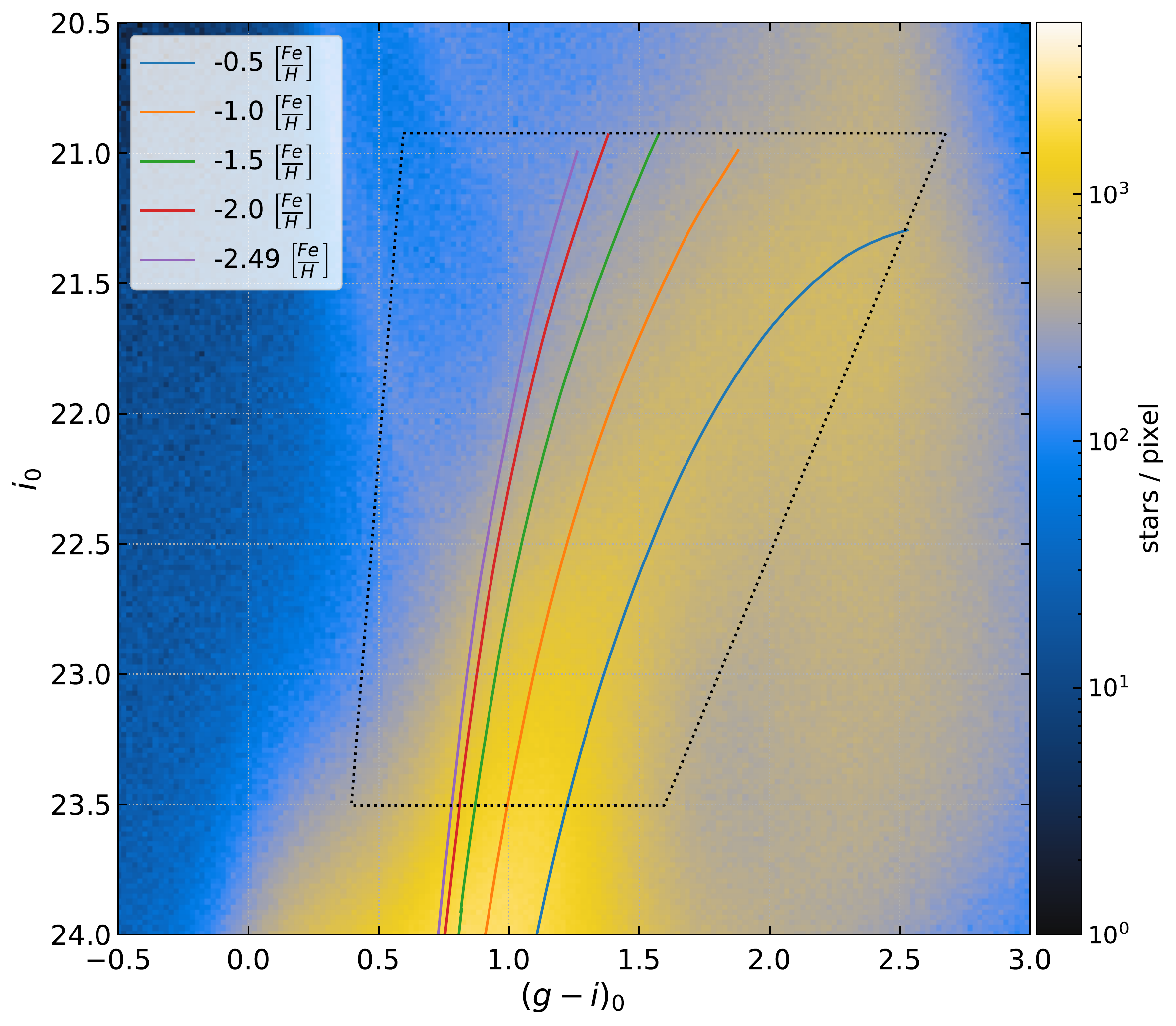}
    \caption{Color-magnitude diagram of all stars in a 150 kpc radius of M31 from the PAndAS survey with a $0.02 \times 0.02$ magnitude pixel size. The selection box boundaries are the same as in \protect\cite{Martin_2013}, where the upper and right box limit is determined by the tip of the red giant branch (the highest reaching isochrone) and the right most reaching isochrone (the reddest star) respectively.}
    \label{fig:RGB_box}
\end{figure}

The observed magnitudes as stated in the PAndAS survey, are $g$ and $i$ and their de-reddened equivalent have the $_0$ subscript, which we omit from now on. In the color magnitude space of all these stars, we selected only a sub-sample corresponding to stars which have color and magnitude as expected from old, low-metallicity stars. This \enquote{selection box}, as depicted in Fig. \ref{fig:RGB_box}, is based on isochrones of a stellar population with a certain age and metallicity. For our purpose, we generated\footnote{We used isochrones from \cite{Dotter:2008mw}, which can be generated at the \href{http://stellar.dartmouth.edu/models/isolf_new.html}{Dartmouth Stellar Evolution Database}.} isochrones for six different age populations (3,5,7,9,11 and 13 Gyr), each having five different metallicities, i.e. -[Fe/H] ratios (0.5, 1.0, 1.5, 2.0 and 2.5). They were adjusted for the M31 distance modulus based on \cite{Conn_2011, Conn_2012}.

The data of the PAndAS survey has holes in some places due to gaps in the coverage, saturated bright stars, or instrumental failures. We reconstructed this missing data by using our \texttt{SMILE} inpainting algorithm.

\begin{figure*}
    \centering
    \includegraphics[width=\textwidth]{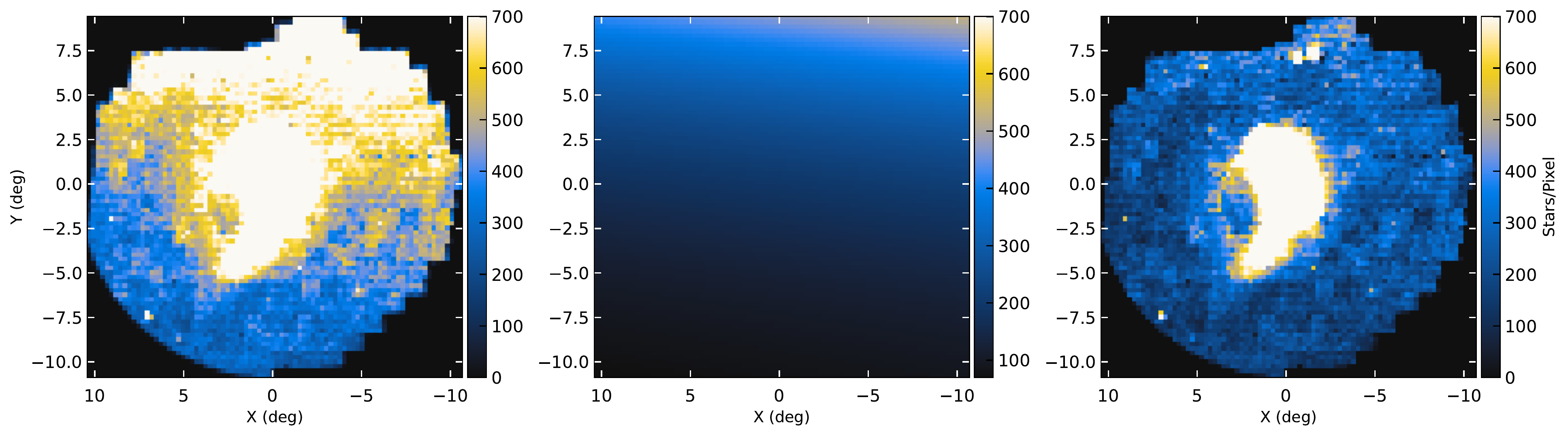}
    \caption{Visualisation of the three stages of the stellar map construction. The coordinates $(X,Y)$ are the equatorial coordinates $(\alpha, \delta)$, projected on a plane tangential to the M31 centered celestial sphere. \textbf{Left:} All stars in our line of sight, where some are part of the Milky Way system. \textbf{Middle:} The model for the stars belonging to the Milky Way system, which are contaminating the sample we are interested in, i.e. all stars of the M31 system. \textbf{Right:} Residual stars after subtracting the contaminant stars, resulting in stars mostly belonging to Andromeda.}
    \label{fig:starSteps_13Gyr}
\end{figure*}

Since we are ultimately interested in the old stars, which are part of the Andromeda system, the stars residing in the halo of the Milky Way are therefore contaminants and have to be removed. The density of these contamination stars, at a specific color and magnitude $(g-i,i)$, is modelled as an exponential increase towards the Galactic plane and center of the Milky Way as

\begin{equation}
    \Sigma_{(g-i,i)} (X,Y) = exp(\alpha_{(g-i,i)} X + \beta_{(g-i,i)} Y + \gamma_{(g-i,i)}),
\end{equation}
where the coordinates $(X,Y)$ are the equatorial coordinates $(\alpha, \delta)$, projected on a plane tangential to the M31 centered celestial sphere. The functions $\alpha, \beta$ and $\gamma$\footnote{These functions were kindly provided to us by the authors of the paper.} are constructed empirically from regions near M31, where contamination of the Milky Way foreground stars is low. For more details regarding these functions see \cite{Martin_2013}. A visualization of the individual steps of the construction for these maps can be seen in Fig. \ref{fig:starSteps_13Gyr}.

This procedure gives us a stellar map for each of our six selected age populations ranging from 3 to 13 Gyr. These are now subtracted from each other in such a way that we get 5 stellar maps containing 3-5, 5-7, 7-11 and 11-13 Gyr population groups. 

Unfortunately the data of this survey is not reliable in the bulge area of M31. This is mostly due to the telescope not being able to resolve individual stars in this bright region of the sky. We therefore follow the common procedure (see e.g. \cite{2014ApJ...787...19M}) of masking the bulge area in these maps and instead use data from the Wide-field Infrared Survey Explorer (WISE) \citep{Wright_2010} at wavelengths of 3.4 (W1) and 4.6 (W2) microns\footnote{We used data available at \href{https://skyview.gsfc.nasa.gov/current/cgi/query.pl}{NASA's SkyView}.}, which was shown to be best suited for tracing stellar light rather than dust (e.g. \cite{Ness_2016}).

\begin{figure*}
    \centering
    \includegraphics[width=\textwidth]{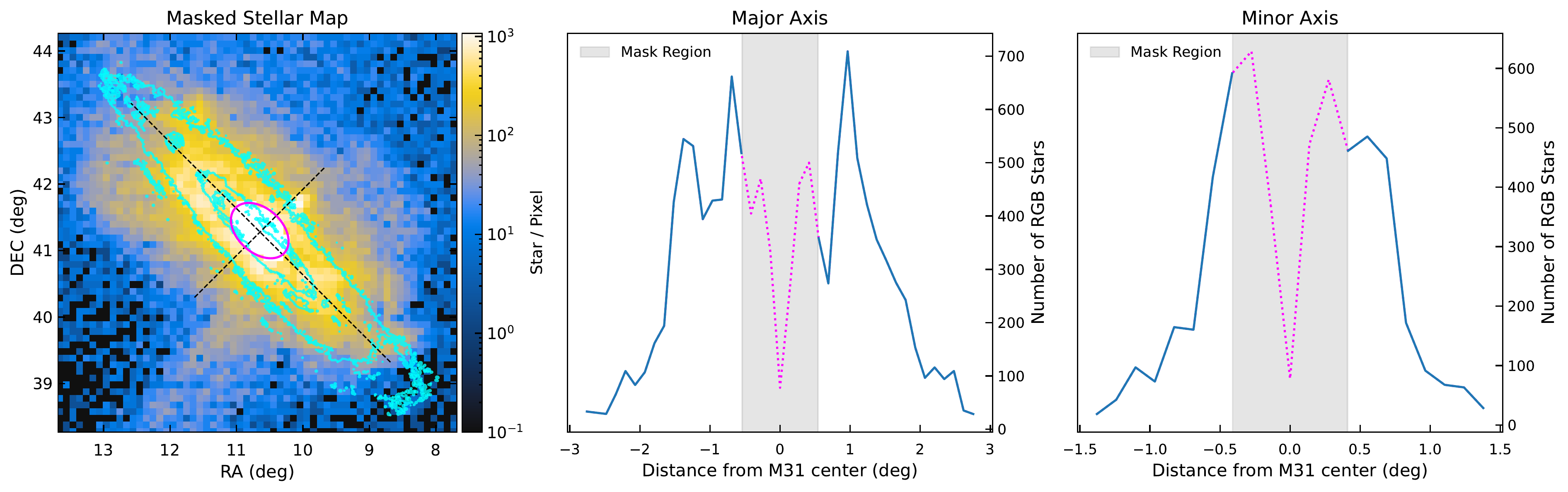}
    \caption{The left panel shows the extent of the mask covering the centralized region, i.e. the bulge of our constructed stellar maps (for the 3-5 Gyr age map in this case). The density of stars along the major and minor axis (the dotted black lines in the left panel) of the elliptical shaped disk of Andromeda is shown in the middle and right panel respectively. The masked region is indicated by the grey shaded area, where we can see the decrease in the number of stars by the dotted magenta lines.}
    \label{fig:starmap_mask}
\end{figure*}

The dimension of this mask is based on a study of the structural parameters of M31 \citep{Courteau_2011}, which contains estimates of the extension of the bulge based on its luminosity profile (see their Fig. 9). We show a close-up of this masked region together with the stellar density profile along the major and minor axis of M31 in Fig. \ref{fig:starmap_mask}. Note that using the raw W1 and W2 templates away from the bulge is not well justified given that these are expected to be heavily contaminated by dust/gas emission from the disk. In contrast to the MW center, we observe the bulge of M31 almost completely unobstructed by its own gaseous disk, due to the large inclination angle of M31. This further supports the assumption that the infrared W1/W2 data in the bulge is mostly produced by the stars.

\begin{figure*}
    \centering
    \begin{subfigure}{\textwidth}
        \centering
        \includegraphics[width=\linewidth]{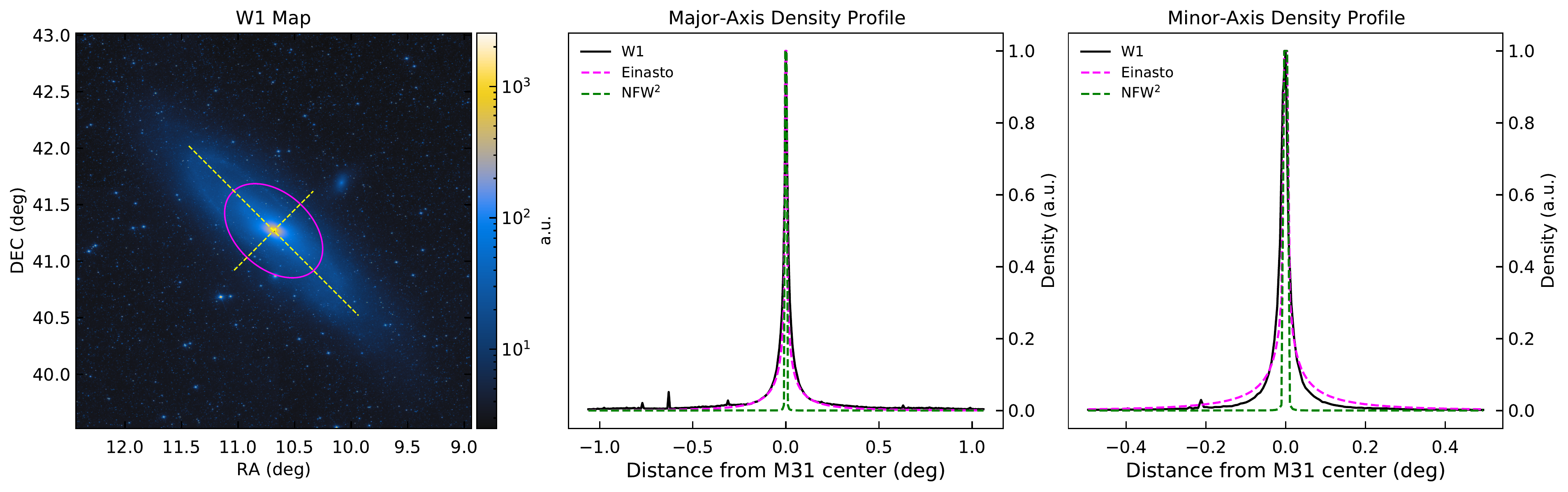}
    \end{subfigure}

    \begin{subfigure}{\textwidth}
        \centering
        \includegraphics[width=\linewidth]{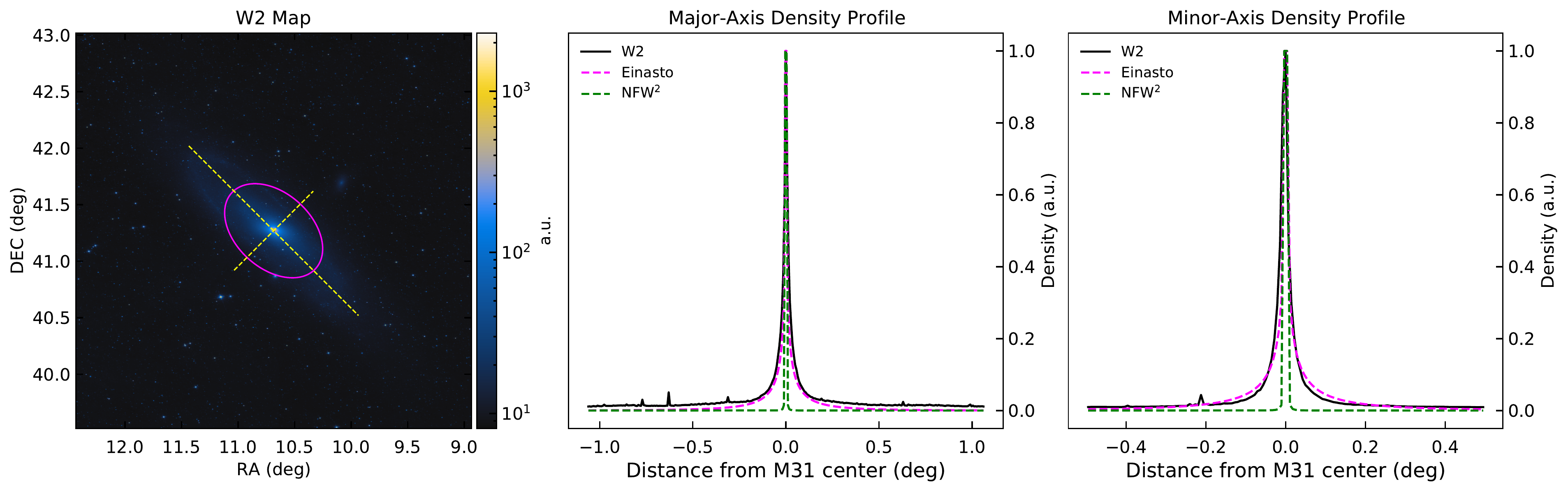}
    \end{subfigure}
    
    \caption{Radial density profile comparison of our W1 and W2 bulge templates along the major and minor axis (the dotted black lines in the left panel) with the bulge Einasto profile used in \protect\cite{PhysRevD.103.083023} and the squared NFW profile. All the panels are shown in arbitrary units (a.u).}
    \label{fig:WISEradialprofiles}
\end{figure*}

One of the key differences in this work compared to previous works, is the use of empirical data over smooth analytical functions for our stellar templates. This guarantees a more sensitive comparison of the MSP to the DM hypotheses in our later analysis. A different approach based on using Einasto density profiles to model the bulge and disk of M31 has been used in e.g. \cite{PhysRevD.103.083023}. We compare their bulge Einasto profile to the radial density profile of our stellar map for W1 and W2 in the top and bottom row of Fig. \ref{fig:WISEradialprofiles}, respectively. The middle and right panel show the normalized density along the major and minor axis, seen as black dotted lines in the left panel, respectively. Additionally, we have added a squared NFW density profile, representing our DM template, demonstrating that it is more concentrated to the innermost regions of M31 compared to the other two profiles, which impacts the results of our morphological analysis.

The combination of these masked stellar maps (each containing a different age group) with either a W1 or W2 bulge acts as our final model for the MSPs hypothesis.

\subsection{Dark Matter Model}
\label{sec:dark_matter}

\begin{figure}
    \centering
    \includegraphics[width=\linewidth]{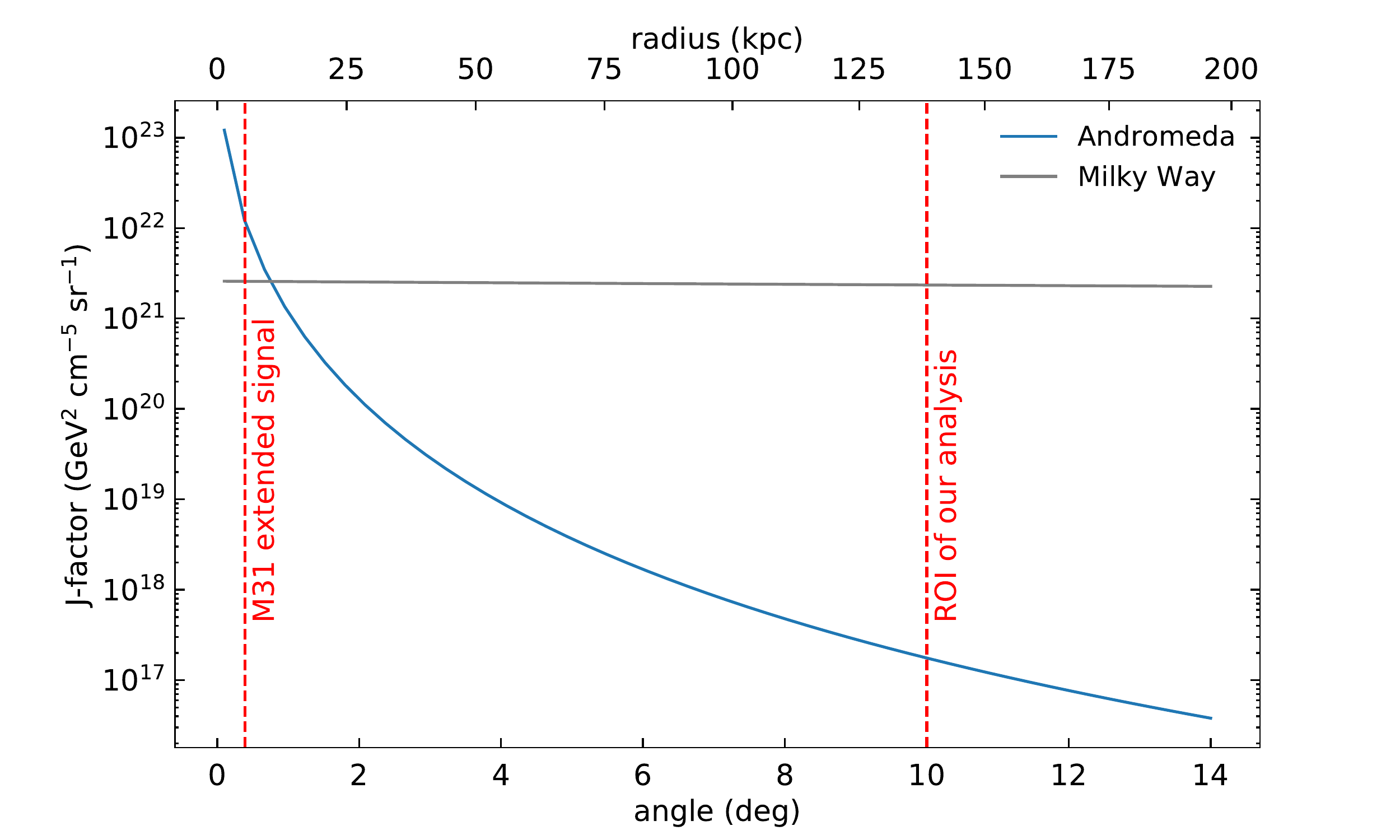}
    \caption{Radial profiles of the J-factor of our DM template for both the M31 and MW halo component, consistent with those of \protect\cite{Karwin:2020tjw}.}
    \label{fig:J_factors_M31_MW}
\end{figure}

Our DM template has to account for both the M31 and MW DM halos, since our line-of-sight extends through both of them. We model the radial density profiles of both halos with the symmetric NFW profile \citep{Navarro:1995iw} given by 

\begin{equation}
    \rho(r) = \frac{\rho_0}{\frac{r}{r_s} (1+\frac{r}{r_s})^2}
    \label{eq:NFW_profile}.
\end{equation}

For the density $\rho_0$ and the scale radius $r_s$ for M31 and the MW we use the parameters from \cite{Karwin:2019jpy}. We begin by expressing the radius from the center of either Andromeda or our Galaxy in terms of the line of sight variable $s$ and the Galactic longitude and latitude coordinates $l$ and $b$ (centered on M31's SIMBAD coordinates in case of the M31 dark matter halo) with the law of cosines as

\begin{equation}
    r = \sqrt{d^2 + s^2 - 2ds \cos(l) \cos(b)}.
\end{equation}

To obtain the value for one pixel of our template, the DM density is squared\footnote{The flux from dark matter annihilation is proportional to the density profile squared, as the interaction requires two particles.} and integrated along the line of sight as

\begin{equation}
    J_p(l,b) = \int_s \rho(r[s,l,b])^2 ds.
\end{equation}

The total value of each pixel $J_p$, also referred to as the J-factor, is then the sum of the contributions of both DM halos. This is repeated for each pixel to obtain the final template. Figure~\ref{fig:J_factors_M31_MW} shows the J-factor profile as a function of  distance from the M31's centre. It is clear that the J-factor contribution from the MW halo becomes dominant for angular distances $\gtrsim 1^\circ$ away from the centroid of M31~\citep{Karwin:2020tjw}. We also display (vertical lines) the size of our ROI in comparison with the extent of the observed gamma-ray excess, which demonstrates the adequacy of our analysis region. Furthermore, we compare our resulting radial J-factor profile with the one obtained by \cite{Karwin:2020tjw} (c.f. their Fig. 6 and our Fig. \ref{fig:J_factors_M31_MW}) and find our values to be consistent with theirs.

\subsection{Fitting Procedure}

Since the flux ratios of the components included in the {\it Fermi} background model were obtained from an all-sky fit, these do not necessarily represent well the flux ratios in a small patch of the sky, such as the 14 by 14 degree region around M31 we have in this work. Another important issue is that this compactified map does not permit a rigorous evaluation of the systematic uncertainties associated to the Galactic diffuse emission model.

In this work, we investigate the impact of each component making up the Galactic diffuse emission. For this, we fit all components of our alternative foreground emission model (FEM), as summarized in Table \ref{tab:model_components}, individually using a bin-by-bin analysis procedure. We run a separate maximum likelihood fit in each of these bins, where we use a simple power-law ($dN/dE =N_0 E^{-\alpha}$) to model the spectra of every source in our ROI. We allowed the normalization $N_0$ of each source to vary, such that they have freedom to absorb any potential excess that is due to Galactic foreground emission. However, the spectral slope $\alpha$ was set to $2$ to stay agnostic to the overall unknown spectral shape. The fits were performed with the \texttt{PyLikelihood} tool, where a total of 79 point sources were present in our ROI. The obtained likelihoods were used to compute the statistical significance of the various hypotheses in $\sigma$ units through equations \ref{eq:likelihood_ratio} and \ref{eq:significance}.

\section{Results}
\label{sec:results}

The contents of the previous sections can be summarized as follows. There are certain challenges we face when dealing with the complex nature of gamma-ray emission from M31. The main challenge is how to disentangle the foreground emission due to the MW from the emission in M31. We do so by constructing tailor-made emission templates for our ROI with the use of state-of-the-art image restoration techniques. The alternative M31-free foreground templates are used in this section to estimate the effects that systematic uncertainties on this component have on the properties of the M31 excess. 

In summary, the foreground components comprising our baseline (also referred to as standard) model for the fits consist of: the {\it Fermi} diffuse emission model described in sect. \ref{sec:fermi_bkg}, the extragalactic diffuse emission model \verb|iso_P8R3_CLEAN_V3_PSF3_v1| for the first three energy bins and \verb|iso_P8R3_CLEAN_V3_v1| for the rest, together with the 79 point sources in our region of interest from the 4FGL-DR2 source catalog. Our alternative foreground model consist of either the H{\it I}4PI survey or the galactocentric H{\it I} and H2 ring templates for the gas-correlated emission, and the models labeled A,B or C for the inverse-Compton emission. The components modeling the extragalactic emission and $\gamma$-ray point sources are the same as in the baseline model. A list of all the templates considered in this work is shown in Table~\ref{tab:model_components}.

\begin{table*}[ht!]
\centering
\begin{tabular}{@{\extracolsep{1pt}} c|c}
\hline
\hline
\textbf{Baseline foreground model components} & \textbf{What $\gamma$-rays does it model?} \\
\hline
{\it Fermi} diffuse emission model & total Galactic diffuse emission \\
\verb|iso_P8R3_CLEAN_V3_PSF3_v1| & extragalactic emission (first 3 energy bins) \\
\verb|iso_P8R3_CLEAN_V3_v1| & extragalactic emission (all other energy bins) \\
4FGL-DR2 & point sources \\
\hline
\hline
\textbf{Alternative foreground model components} & \textbf{What $\gamma$-rays does it model?} \\
\hline
H{\it I}4PI survey & gas-correlated emission \\
galactocentric H{\it I} \& H2 ring templates & gas-correlated emission \\
IC model A,B,C & Galactic inverse-Compton emission \\
\verb|iso_P8R3_CLEAN_V3_PSF3_v1| & extragalactic emission (first 3 energy bins) \\
\verb|iso_P8R3_CLEAN_V3_v1| & extragalactic emission (all other energy bins) \\
4FGL-DR2 & point sources \\
\hline
\hline
\end{tabular}
    \captionsetup{justification=centering}
    \caption{Summary of individual components comprising our baseline (also referred to as standard) and alternative foreground emission model used in our analysis. Each model component is described in section \ref{sec:methods}.}
    \label{tab:model_components}
\end{table*}

\subsection{Spatial Morphology of the M31 excess}

\subsubsection{Phenomenological Models}

We started our analysis runs with the phenomenological models (see Sect. \ref{sec:pheno_models}). We are interested in comparing our findings to previous work on this matter, since we are using the CLEAN class data compared to the more commonly used SOURCE class data, which might influence the results of a morphological analysis significantly.

We tested two cases where the center of the templates is either fixed or free. For the latter we moved the templates through a grid of locations. First, we moved it through a low resolution grid with a stepsize of $0.1$ degrees in (RA,DEC) space. We then followed up with a high resolution grid with a stepsize of $0.01$ degrees in (RA,DEC) space starting from the best-fit spatial position of the low resolution analysis. 
This procedure did not improve the fit much and yielded an almost identical value to the templates with a fixed center, all lying in the range of $5.42\sigma$ to $5.46\sigma$. All the results are summarized in Table \ref{tab:TStable_pheno}.

\begin{figure*}
    \centering
    \begin{subfigure}[t]{0.48\textwidth}
        \centering
        \includegraphics[width=\linewidth]{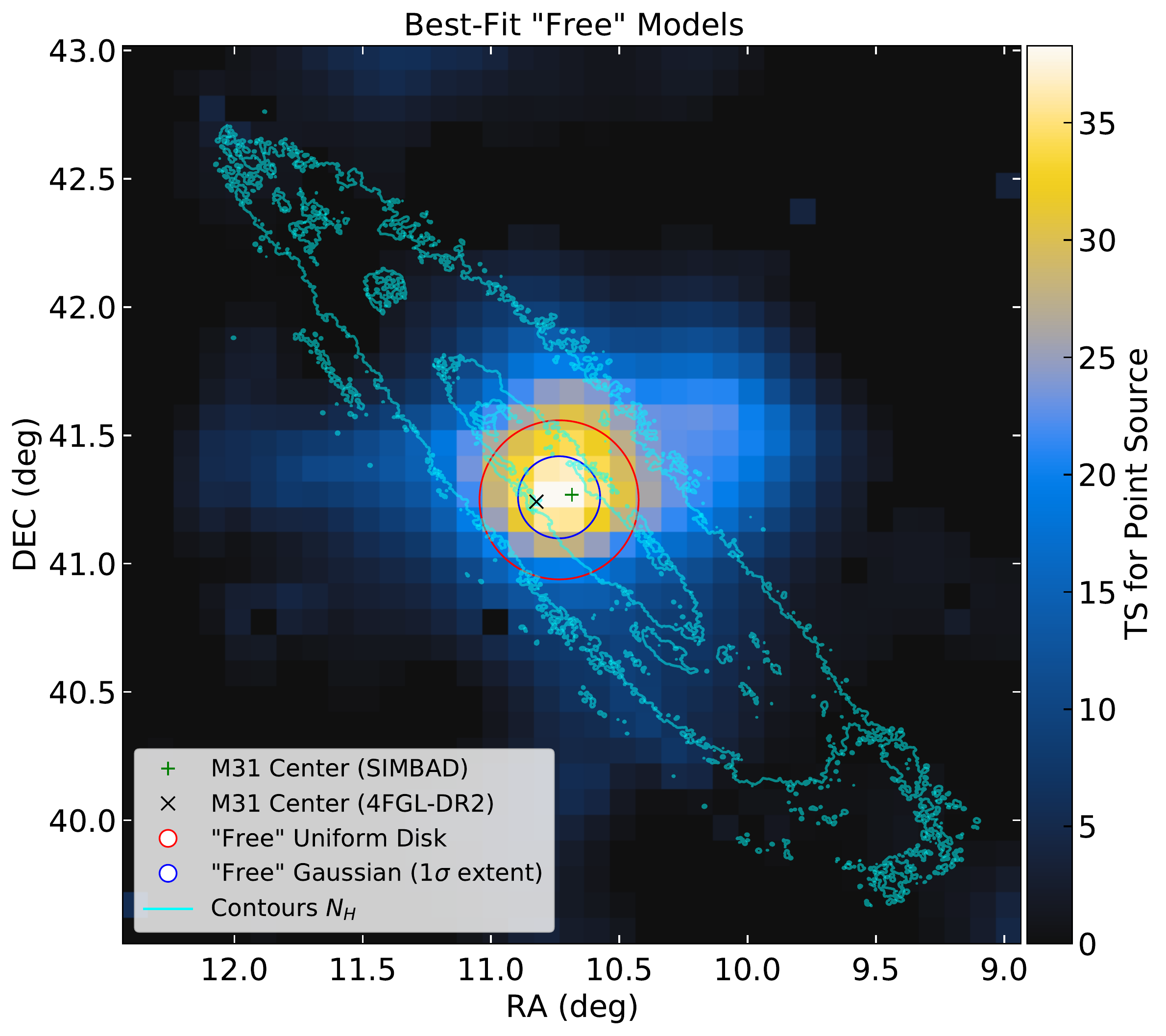}
        \caption{Best-fit position of the phenomenological models for M31, where we scanned over a grid of spatial locations to find the best-fit position.}
        \label{fig:TSmap_with_overlays_FreeModels}
    \end{subfigure}
    \hfill
    \begin{subfigure}[t]{0.48\textwidth}
        \centering
        \includegraphics[width=\linewidth]{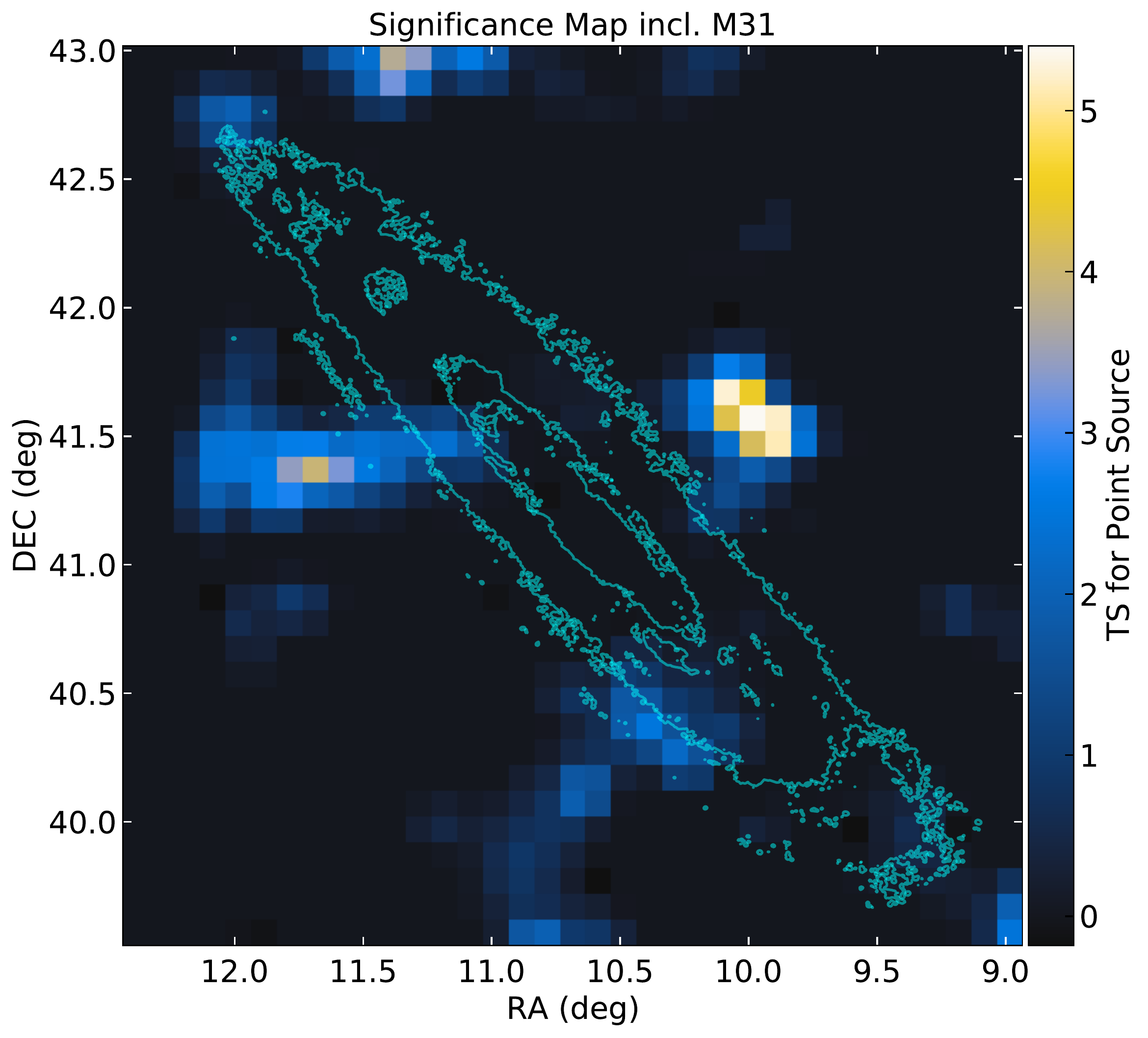}
        \caption{The significance map generated by moving a putative point source through a grid of locations when including the source catalog model for M31. There is no significant unmodelled excess that could be interpreted as a point-like source.}
        \label{fig:TSmap_withM31}
    \end{subfigure}
\end{figure*}

To visualize the extension of these freed models we superimpose them on a significance map of the M31 signal in Fig. \ref{fig:TSmap_with_overlays_FreeModels}. This significance map was obtained by moving a putative point-like source through a spatial coordinate grid and calculating the TS value at that point. The model used to test for the likelihood did not include a model for M31. The resulting map traces the morphology of the gamma-ray emission and is solely meant for visualizing the extent of the emission given our data. 

A quick but important check is to see if there is anything we missed to model in our ROI, such as a point-like source we neglected to include in our background model. To see this, we again generate a significance map but this time we include M31 as modelled in the 4FGL-DR2 catalogue. We do not find any significant unmodelled excess in our ROI as can be seen in Fig. \ref{fig:TSmap_withM31}.

\subsubsection{Physically-Motivated Models}
\label{sec:phys_mot_models}

For our more physically-motivated models (such as e.g. our stellar templates based on empirical observations) we first establish a significance ranking, for which we test each of our templates individually. We present the results of these tests in the upper section of Table \ref{tab:TStable_physics}. We found that our gas and dust templates all yielded similar results in the range from about $2.7$ to $3.0\;\sigma$, showing that the M31 excess is not obviously correlated with emission from gas or dust. We also fitted a 150 MHz radio template, motivated by a study investigating the contribution of MSPs to this radio frequency \citep{Sudoh:2020hyu}, but could not obtain a significantly enough detection for that template with only ${\sim}1.7\;\sigma$. Also, the signal does not seem to come from the disk alone, as our stellar disk templates all have rather low significances between ${\sim}2.2\sigma$ and ${\sim}2.4\sigma$. Most importantly, our results support previous works, finding the signal to be spatially correlated with the bulge of M31. In this ranking, our bulge templates even slightly outrank the DM template, with ${\sim}5.4\sigma$ and ${\sim}5\sigma$ respectively.

\begin{figure*}
    \centering
    \includegraphics[width=\linewidth]{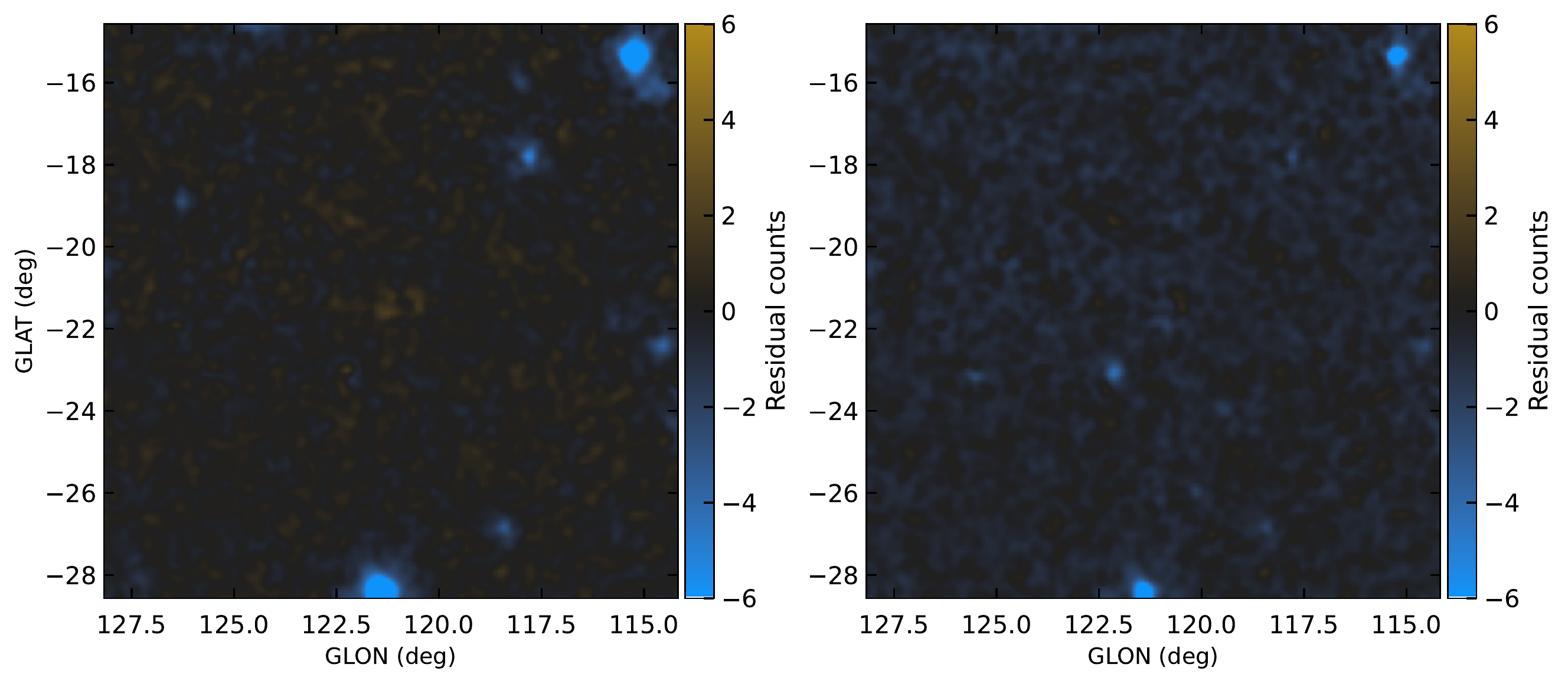}
    \caption{Residual maps when we only fit the {\it Fermi} foreground model (left panel) to the data and if we include our best-fit bulge+disk model (right panel). The maps have been smoothed with a Gaussian kernel covering $1^{\circ}$, approximately corresponding to the PSF of {\it Fermi}-LAT at lower energies.}
    \label{fig:residuals_bulge+disk}
\end{figure*}

Since our stellar map is a two-component model comprised of a stellar disk component incorporating a certain red giant age population together with a component for the bulge based on WISE data, we show the significances for all 10 possible combinations of them. The results are displayed in the lower section of Table \ref{tab:TStable_physics}. We find that all combinations have a very similar significance of around $4.6\;\sigma$. Naturally, we expect the MSPs to be present both in the disk and the bulge of M31. However, due to the weak nature of the excess and the lack of statistics, there is no strong support for both the stellar disk and bulge template together over the bulge template alone. However, we expect this to change with more future data and future $\gamma$-ray missions. We analyze the spectrum of the excess using this combination of templates rather than the bulge alone, since they differ by less than one $\sigma$ and the lack of statistics in the higher energy bins led to fluctuating results, when using the bulge template alone. 
Using the best-fit "disk+bulge" model, we display the residual photon counts map, i.e. the difference of the data to the model, in Fig. \ref{fig:residuals_bulge+disk}. In the left panel, we only fit the {\it Fermi} foreground model to the data, whereas in the right panel we added the disk+bulge model. The inclusion of this model has the effect of smoothing out the residuals, especially in the center of the ROI, where the M31 excess is located.

\begin{table*}[ht!]
\centering
\begin{tabular}{ c c c c c c} 
\hline
\hline
M31 Template & TS & $\sigma$ & R.A. (deg) & Decl. (deg) & Radius (deg)  \\ 
\hline
Point source (fixed) & 41.83 & 4.92 & 10.6847 & 41.2687 & -  \\
Point source (free) & 42.02 & 4.94 & 10.7047 & 41.2787 & -  \\
Uniform disk (fixed center) & 43.53 & 5.42 & 10.6847 & 41.2687 & 0.31  \\
Uniform disk (free center) & 44.04 & 5.46 & 10.7347 & 41.2487 & 0.31  \\
Gaussian disk (fixed center) & 43.46 & 5.42 & 10.6847 & 41.2687 & 0.16  \\
Gaussian disk (free center) & 43.80 & 5.44 & 10.7347 & 41.2587 & 0.16  \\
\hline
\hline
\end{tabular}
\captionsetup{justification=centering}
\caption{Significances for our phenomenological models for M31. The number of degrees of freedom of each template, i.e. the number of parameters left free to vary in the fit ($N_{d.o.f.}$), corresponds to 9.}
\label{tab:TStable_pheno}
\end{table*}

\begin{table*}[ht!]
\centering
\begin{tabular}{ c c c c } 
\hline
\hline
M31 Template(s) & TS & $\sigma$ & $N_{d.o.f.}$ \\ 
\hline
Braun/BraunV2/BraunV3 & 17.72/15.92/16.63 & 2.89/2.66/2.75 & 9 \\
Herschel/Spitzer & 18.47/18.95 & 2.98/3.04 & 9 \\
150 MHz Radio & 9.26 & 1.68 & 9 \\
NFW Dark Matter & 38.42 & 5.00 & 9 \\
W1/W2 Bulge & 43.79/43.58 & 5.44/5.43 & 9 \\
Stellar Disk (3-5 Gyr) & 13.82 & 2.38 & 9 \\
Stellar Disk (5-7 Gyr) & 14.16 & 2.42 & 9 \\
Stellar Disk (7-9 Gyr) & 13.69 & 2.36 & 9 \\
Stellar Disk (9-11 Gyr) & 13.02 & 2.26 & 9 \\
Stellar Disk (11-13 Gyr) & 12.65 & 2.21 & 9 \\
\hline
Stellar Disk (3-5 Gyr) + W1/W2 Bulge & 46.38/45.92 & 4.75/4.72 & 2 $\times$ 9 \\
Stellar Disk (5-7 Gyr) + W1/W2 Bulge & 46.28/45.81 & 4.75/4.71 & 2 $\times$ 9 \\
Stellar Disk (7-9 Gyr) + W1/W2 Bulge & 46.30/45.88 & 4.75/4.72 & 2 $\times$ 9 \\
Stellar Disk (9-11 Gyr) + W1/W2 Bulge & 40.32/45.39 & 4.25/4.67 & 2 $\times$ 9 \\
Stellar Disk (11-13 Gyr) + W1/W2 Bulge & 45.78/45.36 & 4.71/4.67 & 2 $\times$ 9 \\
\hline
\hline
\end{tabular}
\captionsetup{justification=centering}
\caption{Significances for our physically motivated models and the two-component stellar maps. The TS and $\sigma$ values seperated by the slash symbol refer to the Templates seperated by the slash symbol in the same order. E.g. the right TS and $\sigma$ value in the last row corresponds to the Stellar Disk (11-13 Gyr) + W2 Bulge template.}
\label{tab:TStable_physics}
\end{table*}

\subsection{Spectrum of the stellar disk plus bulge model}

\begin{figure}
    \centering
    \includegraphics[width=\linewidth]{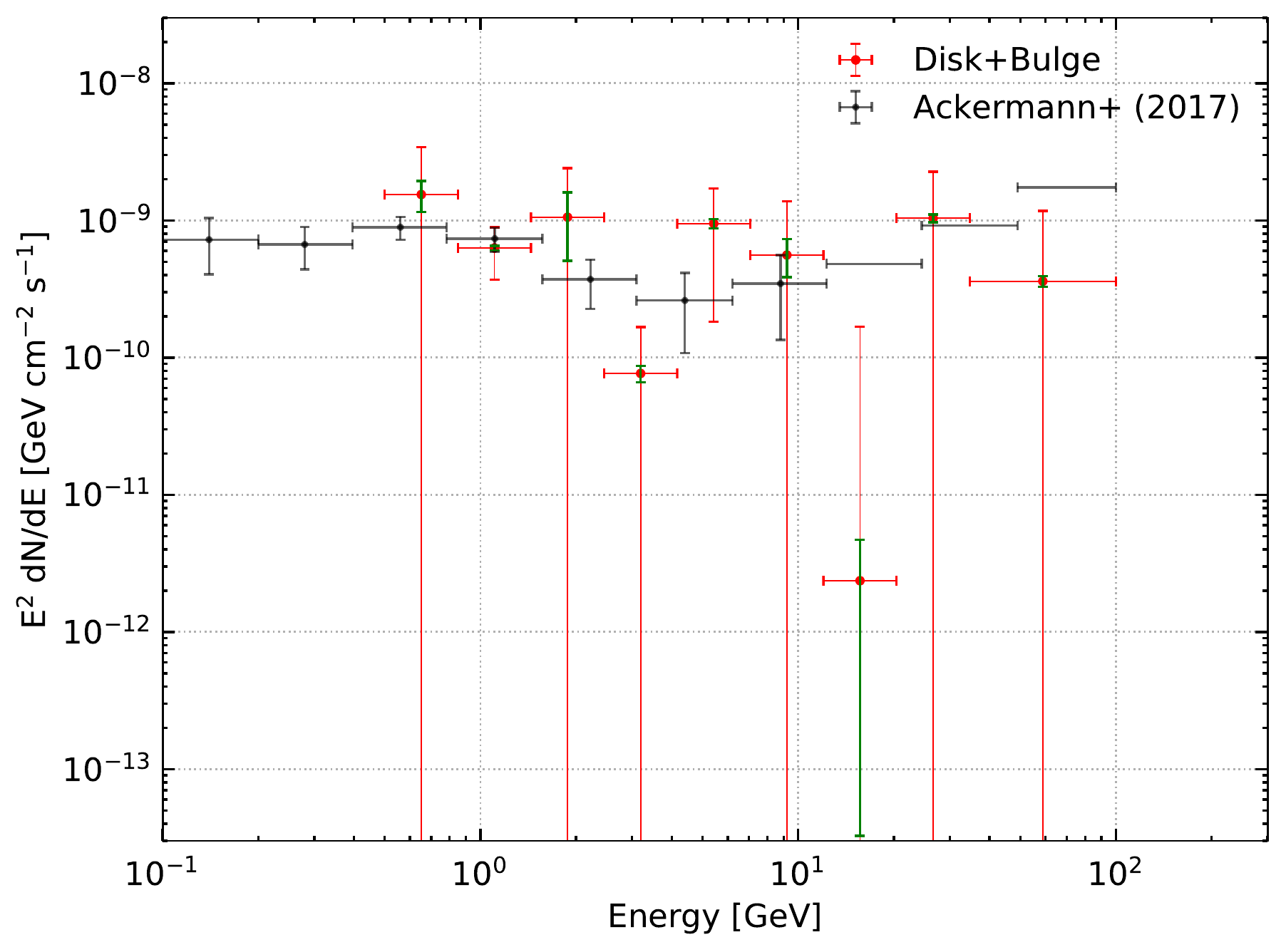}
    \caption{Spectrum of the fit with M31 modelled with the stellar disk (3-5 Gyr) + W1 bulge template, representing a putative MSP population. The red and green error bars represent the statistical and systematical errors, respectively. The flux values when using a uniform disk template of a previous analysis from \citep{Ackermann:2017nya} are shown in black.}
    \label{fig:spectrum_stellarDisk+Bulge}
\end{figure}

We also investigate the spectrum of our best-fit stellar disk + bulge model, i.e. the model for the MSP population in the disk and bulge of M31. The spectrum together with previous results of the {\it Fermi} collaboration is shown in Fig. \ref{fig:spectrum_stellarDisk+Bulge}. The flux values are comparable to the study first documenting the Andromeda excess \cite{Ackermann:2017nya} and to one of the most recent studies \cite{PhysRevD.103.083023} ensuring the validity of our results. 

To accomplish our final goal, which is obtaining the alternative background induced systematic uncertainties for this best-fit model, we adopt the procedure of \cite{Acero:2015prw}. The authors of this study have developed a method to calculate these uncertainties, in which they employ alternative interstellar emission models in addition to their standard (STD) model. Using the notation of \cite{Biteau:2018tmv}, the systematic error is then calculated as

\begin{equation}
    \delta P_{\text{sys}} = \sqrt{ \frac{1}{\sum_{i} \sigma_i^{-2}} \sum_{i} \sigma_i^{-2} (P_{\text{STD}} - P_i)^2 }.
    \label{eq:sys_uncertainty}
\end{equation}

In this equation, $P_{\text{STD}}$ is the value for which we want to find the systematic error of, i.e. the flux value when using the standard background model. The flux values and their statistical errors when using the alternative background models are $P_i$ and $\sigma_i$ respectively. In our case, we used the {\it Fermi} background as our standard model and the 24 FEMs as our alternatives. The resulting systematic uncertainties are shown as the green error bars in Fig. \ref{fig:spectrum_stellarDisk+Bulge}. For our case the statistical errors dominate over the systematic uncertainties. The reason for these large errors in our case as opposed to the much smaller errors in the recovered spectrum in the original study of the M31 excess \citep{Ackermann:2017nya} is the following. In that study, the individual normalizations of the point sources in the bin-by-bin analysis were kept close to the values obtained in their broadband (i.e. over the whole energy range) analysis. We stayed agnostic to the overall shape of the spectrum by only performing a bin-by-bin analysis and keeping the normalizations of all sources in our ROI completely free, since we wanted to understand how this would effect the results. It turns out that the statistics are too low to make a definite statement about this. It is  interesting however, that the excess survives all parts of our analysis with our rigorous and agnostic treatment of the data.

\subsection{Implications for the DM interpretation}

We have tested how strong the DM hypothesis is and if it can survive the inclusion of our more physically-motivated models. We do this with a template-nesting approach similar to the one used in \cite{DiMauro:2019frs}. Even if we assume that parts of the emission are due to annihilating DM, there is still the emission of astrophysical nature, which has to be accounted for. We therefore successively include templates for emission by MSPs in the disk and bulge of M31, as well as a template for gas- and dust-correlated emission into the FEM. The chosen templates are based on the significance ranking established in section \ref{sec:phys_mot_models} and summarized in Table \ref{tab:TStable_physics}.

\begin{figure}
    \centering
    \includegraphics[width=\linewidth]{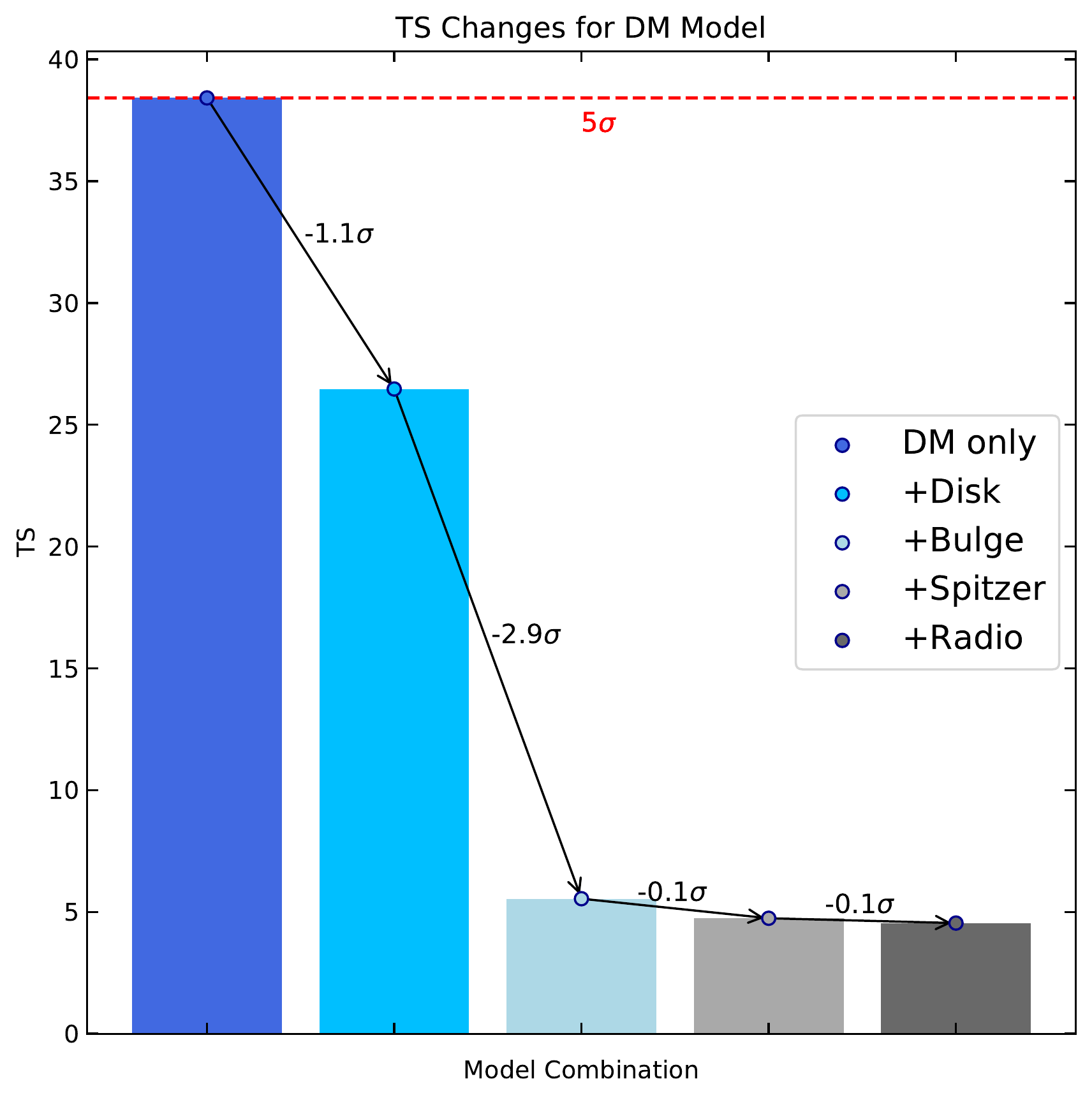}
    \caption{The decrease in significance of detection of the DM template, when we one-by-one include a stellar disk, a bulge, a gas and a radio template.}
    \label{fig:DM_TS_decrease}
\end{figure}

The resulting changes in the TS value of the DM hypothesis can be seen in Fig. \ref{fig:DM_TS_decrease}. If we only use the {\it Fermi} FEM and no additional components are included, the TS value reaches a ${\sim}5\;\sigma$ significance, but quickly falls off significantly (${\sim}1\;\sigma$) as we account for the astrophysical emission from putative MSPs in the bulge and disk of M31  as well as emission from gas and dust. In summary, there is no support for DM related emission once the stellar templates are included.

\subsection{Luminosity}
Using the spectrum of the best-fit model, we calculate the $\gamma$-ray luminosity of M31 as follows. We calculate the flux for each bin by assuming a power law anchored to $E_0 = 1000 \, \mathrm{MeV}$ as
\begin{equation}
    F_\mathrm{{bin_i}} = N_0 \int_{E_{\mathrm{bin_i, min}}}^{E_{\mathrm{bin_i, max}}} \frac{dN}{dE} dE
    = \frac{L}{4 \pi d_{\mathrm{M31}}^2} \int_{E_{\mathrm{bin_i, min}}}^{E_{\mathrm{bin_i, max}}} \left( \frac{E}{E_0} \right)^{-\gamma} dE, 
\end{equation}
where we have expressed the normalisation constant $N_0$ in terms of the luminosity $L$, the distance to M31 $d_{\mathrm{M31}}$, and the photon energy flux above 500 MeV $G_{500}$, which is obtained with
\begin{equation}
    G_{500} = \int_{E_{\mathrm{min}}}^{E_{\mathrm{max}}} E \left( \frac{E}{E_0} \right)^{-\gamma} dE,
\end{equation}
with the limits of the whole energy range considered in this work of $E_{\mathrm{min}} = 500 \, \mathrm{MeV}$ and $E_{\mathrm{max}} = 10^5 \, \mathrm{MeV}$. We can then optimize the $\chi^2$ quantity 
\begin{equation}
    \chi^2 = \sum_i \frac{(F_\mathrm{bin_i} - F_\mathrm{bin_i, mod})^2}{F_\mathrm{bin_i, err}^2}, 
\end{equation}
where $F_\mathrm{bin_i, mod}$ are the flux values of the model for each bin and $F_\mathrm{bin_i, err}$ its associated statistical errors. The optimization was performed with the \textsc{minuit} algorithm \citep{10.1093/comjnl/7.4.308}, available in the \texttt{iminuit} package\footnote{https://iminuit.readthedocs.io/en/stable/}. With this we obtain a total luminosity for M31 of
\begin{equation}
    L_{\rm \gamma,M31} = (1.6\pm0.7) \cdot 10^{38} \, {\rm erg/s} \, ,
\end{equation}
a luminosity for the stellar bulge of 
\begin{equation}
    L_{\rm \gamma,M31,B} = (1.1\pm0.2) \cdot 10^{38} \, {\rm erg/s} \, ,
\end{equation}
and a 95\% C.L. upper limit for the stellar disk of
\begin{equation}
    L_{\rm \gamma,M31,D} < 3.2 \cdot 10^{38} \, {\rm erg/s} \, .
\end{equation}
Note that the disk luminosity is only an upper limit since it was not significantly detected while, on the other hand, a bulge signal is detected at better than $5 \sigma$ confidence.

In order to gauge the  $\gamma$-ray emissivity per unit mass of the M31 stellar structures we note that the total mass of the galaxy \citep{Tamm:2012} is $(1.0-1.5) \times  10^{11} \, M_\odot$ which we shall understand below as
\begin{equation}
    M_{\rm \star,M31} =  (1.25 \pm 0.25) \times 10^{11} \ M_\odot \, ,
\end{equation}
of which $\sim 56 \%$ is the disk
\begin{equation}
    M_{\rm \star,M31,D} =  (7.0 \pm 1.4) \times 10^{10} \ M_\odot \, ,
\end{equation}
with the remainder in the bulge 
(+ halo, using the notation of \citealt{Tamm:2012}). A recent, more tightly-constrained determination \citep{BlanaDiaz2018} for the stellar mass of the M31 bulge  (assuming an NFW form for its DM distribution) is that it  is
\begin{equation}
    M_{\rm \star,M31,B} = (3.0^{+0.1}_{-0.2}) \times  10^{10} \, M_\odot  \, .
\end{equation}
From these measurements we infer the following central values (upper limit in the case of the disk) for the $\gamma$-ray emissivity per unit stellar mass
\begin{eqnarray}
    L_{\gamma,\star,M31} & = & (1.3 \pm 0.6) \cdot 10^{27} \, {\rm erg/s} \\
     L_{\gamma,\star,M31,B} & = & (3.6 \pm 0.3) \cdot 10^{27} {\rm erg/s}\\
      L_{\gamma,\star,M31,D} &<&  5.2 \cdot 10^{27} {\rm erg/s} \, .
\end{eqnarray}

\subsection{Millisecond Pulsar Interpretation}

Here we consider whether MSPs belonging to M31 can supply its detected $\gamma$-ray signal.
The stellar population of M31 is predominantly old with its bulge and disk populations of similar mean age \citep{Tamm:2012}. On the basis of the data presented by \citet{Williams2017}, we determine a mass-weighted mean stellar age of 10-11 Gyr across the entire galaxy. At such an age, the recent binary population synthesis modelling by \citet{2021arXiv210600222G} suggests that the MSP population of M31 numbers around $10^6$, and the spin-down power liberated by magnetic braking of these pulsars is $\sim 5 \times 10^{28}$ erg/s$M_\odot$; this is easily sufficient to sustain the observed $\gamma$-ray luminosity\footnote{Note that \citet{2021arXiv210600222G} model a specific channel for the production of MSPs, viz.~the accretion induced collapse (AIC) of O-Ne white dwarfs. Were other channels to the production of MSPs  in operation in this old stellar population (e.g., `recycling' of old, core-collapse neutron stars due to accretion from a binary companion), the spin down power available would be even larger. AIC is, however, perhaps singularly good at delivering new MSPs at time separations of many Gyr post star formation; see \citep{Ruiter2019}.}.

Despite the energetics being easily met, two question marks hang over any putative identification of the M31 $\gamma$-rays  with an MSP signal: i) the  rather hard spectrum shown in the SED plot (\autoref{fig:spectrum_stellarDisk+Bulge})
does not resemble the classical $\sim 2$ GeV bump spectrum of $\gamma$-ray MSPs, pulsars, or, indeed, the GCE; and ii) despite the strong detection of the bulge, the disk signal is only marginally detected. Given, as already noted, these stellar populations are characterised by similar mean ages, why does the disk not produce an easily-detectable $\gamma$-ray signal from its MSPs given its stellar mass actually exceeds that of the bulge?

Of course, we should grant at the outset that, formally, neither of these points is a show-stopper
given the scale of present uncertainties/errors in our analysis (i.e., the large error bars on the $\gamma$-ray SED we have measured would accommodate a bump-like signal and the 95\% upper limit on the disk $\gamma$-ray efficiency is clearly compatible with the measured efficiency of the bulge). Nevertheless, in anticipation of tighter constraints, we point out here that there is a natural scenario involving MSPs that would accommodate both a hard SED from the M31 bulge and a small, perhaps undetectable, signal from the disk. This, namely, is that the overall bulge signal is dominated not by the aggregate, `prompt' curvature radiation from the magnetospheres of all the individual (albeit unresolved) MSPs, but rather by the inverse-Compton up-scattering of ambient photons by the cosmic ray electrons and positrons launched by these MSPs into the bulge ISM \citep[cf.][]{Petrovic2015,Ackermann:2017nya,Song2021}. Such emission can easily reproduce the observed hard spectrum up to 10s of GeV \citep[e.g.,][]{2021arXiv210600222G}. 
A natural explanation, then, of why we do not detect (or only marginally detect) such IC emission from the disk is that, in this environment, the ratio $u_B/u_{\rm ISRF}$ of the magnetic field to the local interstellar radiation field energy densities is significantly smaller than in the bulge (see below). Indeed, consistent with this, the high concentration of stars in the M31 bulge guarantees that the ISRF should reach a peak in this region. 

To put some numbers to this,
on the basis of the stellar luminosity we\footnote{See Roth, Krumholz, Crocker et al., forthcoming} calculate a total radiation field energy density in the M31 bulge of 12 eV cm$^{-3}$ and a 0.7 
eV cm$^{-3}$ in the disk at the approximate radius of the star forming ring (10 kpc).
For the magnetic field,
according to radio continuum studies
(and making the equipartition assumption)
\citep{Hoernes1998,Giessubel2014}, the  field amplitude in M31 
reaches a peak at a radius of $800-1000$ pc into the bulge; the measured value here, 
$19 \pm 3 \ \mu$G, thus defines an upper limit
to the ISM magnetic field in the bulge region
from where the measured $\gamma$-ray signal emerges.
On the other hand, the magnetic field amplitude at $\sim 10$ kpc radius coincident with the star-forming ring is ``3-4 times'' \citep{Hoernes1998} smaller than  this.
Overall we thus infer
\begin{equation}
\left. \frac{u_B}{u_{\rm ISRF}} \right|_{\rm M31,B}  <  0.8 \qquad , \qquad
\left. \frac{u_B}{u_{\rm ISRF}} \right|_{\rm M31,D}  \sim  1.1 \, ,
\end{equation}
which implies a transition from IC loss dominance in the bulge to synchrotron loss dominance in the disk, consistent with our scenario (albeit poorly constrained given the uncertainties).
An additional, practical consideration is that a postulated $\gamma$-ray
signal related to the disk stars would be distributed over a significantly larger solid angle than the bulge signal and may be soaked up in one or other of the extended foreground templates.
An obvious extension to our analysis here would be to introduce a simultaneous spectro-morphological treatment of the non-thermal radio data covering M31 \citep[e.g.,][]{Tabatabaei2013} in addition to the $\gamma$-ray data; such is beyond the scope of this paper.

Overall, our results are thus consistent with previous findings that the M31 signal, despite the hard spectrum (assuming the central values of the data points are close to correct) and despite the weak or even (consistent with) null detection of the massive stellar disk, is completely consistent with a dominantly MSP origin.

\section{Discussion and Conclusions}
\label{sec:conclusions}

In this work, we have performed a re-analysis of 10 years of {\it Fermi}-LAT data in the GeV energy range from the M31 region. We have employed novel image reconstruction techniques to derive a multitude of alternative foreground models, giving us a robust framework for the evaluation of the systematic uncertainties in the Galactic diffuse gamma-ray emission models. We have concentrated on the morphology of the emission of M31 by constructing empirical templates, which best represent the stellar population in the bulge and disk of M31. We have opted for this approach instead of using analytical functions like the Einasto profile to avoid possible morphological degeneracies with the DM distribution -- as it is popularly modelled by density distributions similar to Einasto profiles. 

The main results from our analyses, which involved the use of multiple foreground models, can be summarized as follows:
\begin{itemize}
    
    \item In this work we used {\it Fermi}-LAT photon events with improved directional reconstruction at low energies (PSF3 for the first three energy bins).  One of our main goals was to have a better handle on the spatial morphology of the signal, whereas at low energies the angular resolution of the LAT instrument worsens. This implied a trade off between angular resolution and statistical power of the recovered signal for M31. Additionally, we varied the fluxes of all sources included in our region of interest. These two factors combined implied larger (statistical) error bars for the energy spectrum of M31, compared to previous studies. Based on this spectrum we could determine a $\gamma$-ray luminosity of $L_{\rm \gamma,M31} = (1.6\pm0.7) \cdot 10^{38} \, {\rm erg/s}$, which lies within the range of values compatible with the results in e.g., \cite{Ackermann:2017nya}.

    \item The M31 excess was robustly detected despite the large set of alternative foreground models used in our analysis. By testing models for the signal (e.g.,  stellar templates) with different approaches of modelling the $\gamma$-ray foreground emission, we could estimate the systematic uncertainties related to the modeling of these complex diffuse emissions. We conclude that it is unlikely that the M31 excess is caused by statistical fluctuations of the Galactic foreground. We also found that that this excess emission appears localized within the bulge and disk regions of M31.

    \item We have tested stellar templates specifically constructed for this region of the sky to model the emission of a putative population of MSPs in the bulge and disk of M31. We have found these to be as strong as the widely used phenomenological models -- a circular disk template with a uniform or Gaussian brightness profile -- both sitting at the $5.4\;\sigma$ level. With future observations, we expect that more accurate stellar templates will be able to outperform the more simple phenomenological models.
    
    \item Our findings do not support the star formation scenario, in which the regions rich in gas and dust contribute to the gamma-ray emission, since we do not detect the Spitzer and Herschel templates. As we also do not detect any of the hydrogen maps (Braun, BraunV2 and BraunV3) individually, the scenario where the main contribution comes from hadronic-only processes is unlikely as well. 
    These findings are consistent with previous works (e.g. \cite{Ackermann:2017nya}) and can be linked to the properties of M31, specifically to the fact that the star formation rate of M31 was found to be about 10 times smaller compared to that of the MW \citep{Ford:2013}, which can lead to a decreased contribution to the emission from the disk.
    
    \item We have tested whether the DM hypothesis  survives the inclusion of the stellar  templates. By nesting the stellar and DM templates during the fit, we find that the significance of the DM template drops to the $1\;\sigma$ level. We therefore come to a similar conclusion as in related studies using this methodology (e.g. in \cite{DiMauro:2019frs,PhysRevD.103.083023} for M31 and e.g. in \cite{Macias:2019omb} for the MW): There is no significant leftover $\gamma$-ray emission which can be attributed to annihilating DM in the center of M31, when maps of stellar mass are included in the fit.
    
    \item We used the results of the binary population synthesis modelling by \cite{2021arXiv210600222G} to determine whether the MSPs theory  could explain the observed luminosity per stellar mass, hard spectrum, and high stellar bulge-to-disk flux ratio. We found that (i) the observed energetics are easily met if M31 hosts about $10^6$ unresolved pulsars with an average spin-down luminosity of $\sim 5 \times 10^{28}$ erg/s$M_\odot$, and (ii) since the electrons injected by the MSPs lose energy more efficiently in the bulge than in the disk through IC, it is reasonable to conclude that both the hard spectrum and high bulge-to-disk ratio could be explained by an MSPs IC emission scenario. A potential detection of a high-energy tail (at TeV energies) in the M31 spectrum would provide strong support for such a scenario.  In future work, we will perform a sensitivity analysis similar to those by e.g., \cite{Macias:2021boz, Song:2019nrx} to investigate the capabilities of the upcoming Cherenkov Telescope Array to a tentative high-energy tail from MSPs in M31. 
\end{itemize}

\section*{Acknowledgements}
We would like to thank Giuseppe Puglisi for all the help he provided with his inpainting package and Nicolas Martin for providing results from his previous works. We would also like to thank Ariane Dekker, Ebo Peerbooms, Sahaja Kanuri and Manuel Loparco for fruitful discussions and their support. OM was supported by the GRAPPA Prize Fellowship. The work of SA was supported by JSPS/MEXT KAKENHI Grant Numbers JP17H04836, JP20H05850, and JP20H05861. The work of SH is supported by the US Department of Energy under the award number DE-SC0020262 and NSF Grant numbers AST-1908960 and PHY-1914409. This work was supported by World Premier International Research Center Initiative (WPI Initiative), MEXT, Japan.

\section*{Data Availability}


\bibliographystyle{mnras}
\bibliography{bib} 

\appendix

\section{Statistical Framework} \label{sec:stat_frame}

To test for the significance of the model under consideration, we used a bin-by-bin analysis, where we originally divided the data into 10 logarithmically spaced energy bins, but later combined the last two into one to compensate for the lack of statistical power in these bins, which comes from the low photon count in this high energy range. For each bin we performed an independent maximum likelihood fit and the test statistic for each bin is calculated as the likelihood ratio

\begin{equation}
    TS_{\text{bin}} = -2 \, \text{ln} \, \frac{\pazocal{L}_{\text{null}}(\boldsymbol{\hat{\theta}}_{\text{BKG}})}{\pazocal{L}_{\text{alt}}(\boldsymbol{\hat{\theta}}_{\text{M31}},\boldsymbol{\hat{\theta}}_{\text{BKG}})},
    \label{eq:likelihood_ratio}
\end{equation}

where $\pazocal{L}_{\text{null}}$ is the sum of the poisson likelihoods (one for each bin) for the background only and $\pazocal{L}_{\text{alt}}$ is for the hypothesis, where we included the model for M31 on top of the background with $\boldsymbol{\hat{\theta}}_{\text{M31}}$ and $\boldsymbol{\hat{\theta}}_{\text{BKG}}$ representing the parameters for the M31 model and background respectively. For the background we include all sources from the 4FGL-DR2 catalog in our 10 degree region of interest together with a model for the interstellar diffuse gamma-ray emission and the isotropic gamma-ray background. These background models are either the ones from the {\it Fermi} collaboration or our FEMs.

In \cite{Wilks:1938dza} it was shown, that $TS_{\text{bin}}$ is distributed as $\chi^2_k$ in the null hypothesis. The subscript $k$ refers to the independent random variables left free to vary, which are normally called degrees of freedom. This theorem (known as Wilk's theorem) holds as long as the parameters can live in the whole parameter space, but since our parameters live in $\mathbb{R}^+$\footnote{In our analysis we leave the normalization of the templates free to vary in the fit, but it can not be a negative value.}, this theorem no longer holds. In this case we have to use the Chernoff Theorem from \cite{Chernoff:1954eli}, where they showed that for one degree of freedom the TS distribution will then be a mixture of $\chi^2$ and the Dirac delta function $\delta$ as

\begin{equation}
    p(TS) = 2^{-1} (\delta(TS) + \chi^2(TS)).
\end{equation}

This equation implies the following: Under the null hypothesis, 50\% of the time the evaluated amplitude (our parameter left free to vary in the fit) will be negative, in which case the TS will be assigned through the Delta function, and the other 50\% of the time the amplitude is non-negative and the TS distribution will follow from Wilk's theorem. As shown in case 9 of \cite{SelfLiang_Case9}, this holds for any number of degrees of freedom $n$, such that

\begin{equation}
    p(TS) = 2^{-n} \left( \delta(TS) + \sum_{i=0}^n \binom{n}{i} \chi_i^2 (TS) \right).
\end{equation}

As before, the delta function ensures that the values are non-negative, whereas the binomial coefficient $\binom{n}{i}$ accounts for all possible combinations of non-negative amplitudes, each following a $\chi^2$ distribution, and the overall factor of $2^{-n}$ represents all possible arrangements of $n$ energy bins, all having non-negative values. From this, the significance in units of the standard deviation $\sigma$ can then be calculated as

\begin{equation}
    \sigma \equiv \sqrt{ \text{InverseCDF}(\chi_k^2,CDF[p(TS),\widehat{TS}]) },
    \label{eq:significance}
\end{equation}

where (Inverse)CDF is the (inverse) cumulative distribution function and the observed TS value is denoted by $\widehat{TS}$, adopting the notation of \cite{Macias:2019omb}.


\bsp  
\label{lastpage}
\end{document}